\documentclass[aps,pre,reprint,groupedaddress,longbibliography,floatfix]{revtex4-2}

\usepackage{amsmath}
\usepackage{amssymb}
\usepackage{graphicx,psfrag}
\usepackage{subfig}
\usepackage{verbatim}
\usepackage{color}
\usepackage{tabularx}
\usepackage{booktabs}
\usepackage{array}

\usepackage{bm}

\usepackage{comment}

\usepackage{soul}
\usepackage{dirtytalk}
\usepackage{float}

\usepackage{booktabs}
\usepackage{multirow}

\usepackage{calc}
\usepackage{appendix}
\graphicspath{{images/}}

\usepackage{rotating}

\begin{document}

\title{Shaping the equation of state to improve numerical accuracy and stability of the pseudopotential lattice Boltzmann method}
    
    \author{Luiz Eduardo Czelusniak}
    \email{luiz.czelusniak@usp.br}
    
    \author{Vin\'icius Pessoa Mapelli}
    \email{vinicius.mapelli@usp.br}
    
    \author{Luben Cabezas Gómez}
    \email{lubencg@sc.usp.br}
     \address{Heat Transfer Research Group,  Department of Mechanical Engineering, Engineering School of S\~ao 
    Carlos, University of S\~ao Paulo, EESC-USP, S\~ao Carlos, S\~ao Paulo, Brazil}
   
    \author{Alexander J. Wagner}
   \email{alexander.wagner@ndsu.edu}
    \address{Department of Physics, North Dakota State University, Fargo, North Dakota 58108, USA}
     
     \begin{abstract}
     Recently it was discovered that altering the shape of the meta stable and unstable branches of an equation of state (EOS) can greatly improve the numerical accuracy of liquid and gas densities in the pseudopotential method.
     Inspired by this approach we develop an improved approach that is benchmarked 
     for both equilibrium and non-equilibrium situations. We show here that the original approach reduces the method stability in non-equilibrium situations. 
     Here we propose a new procedure to replace the metastable and unstable  regions of these EOS by alternative functions. Our approach does not affects the coexistence densities or the speed of sound of the liquid phase while maintaining continuity of the sound speed in the pressure-density curve. Using this approach we were able to reduce the relative error of the planar interface vapor density compared to the thermodynamic consistent value by increasing the vapor phase sound speed.
     To allow for the benchmarking of dynamic results we also developed a finite difference method (FD) that solves the same macroscopic conservation equation as the pseudopotential lattice Boltzmann method (LBM). With this FD scheme we are able to perform mesh refinement and obtain reference solutions for the dynamic tests. We observed excellent agreement between the FD solutions and our proposed scheme.
     We also performed a detailed study of the stability of the methods using simulations of a droplet impacting on a liquid film for reduced temperatures down to 0.35 with Reynolds number of 300. Our approach remains stable for a density ratio up to $3.38\cdot10^{4}$. 
    \end{abstract}

   \maketitle

\section{Introduction}
\label{sec:intro}

The application of the lattice Boltzmann method (LBM) \cite{chen1998lattice,kruger2017lattice} 
to fluid dynamic simulation has gained much attention in the scientific literature.
Differently from standard numerical methods based on a direct discretization of the
conservation equations, the LBM is based on a discretized form of the Boltzmann transport equation
known as the lattice Boltzmann equation (LBE) \cite{shan1998discretization}.
There are also, many extensions of the LBM to allow its application
to simulate phase change phenomena and multiphase flows \cite{gunstensen1991lattice,swift1996lattice,luo1998unified,shan1993lattice}. 
One of the most popular is the pseudopotential method, which was proposed by Shan and Chen
\cite{shan1993lattice,shan1994simulation}.
The authors proposed a short-range interaction force that could maintain different phases in equilibrium.

The original model proposed by Shan and Chen \cite{shan1993lattice} lacked a direct connection with thermodynamics that underlay the Free energy approaches for phase separation in lattice Boltzmann \cite{swift1995lattice,swift1996lattice,wagner2006thermodynamic}. There have been significant efforts reported in the literature to improve this shortcoming of the original pseudopotential lattice Boltzmann method. 
After proposing the pseudopotential method \citet{shan1993lattice}, tried to predict the equilibrium phase densities induced by their interaction force in a planar interface problem \cite{shan1994simulation}. 
It was observed that these densities did not match with the ones given by thermodynamic theory, unless a particular choice of equation of state is made. The thermodynamic consistent densities are obtained by applying the Maxwell rule \cite{callen1998thermodynamics}. Also, it was not possible to adjust the surface tension independently of the equation of state (EOS). Modifications in the forcing scheme were proposed to allow a better comparison with the Maxwell rule and independent adjustment of surface tension \cite{li2012forcing,li2013lattice,li2013achieving,lycett2015improved}.

A different strategy used to increase numerical stability 
in the pseudopotential method
that does not involve the numerical stencil itself was investigated by some authors in the 
literature. A simple approach to improve the stability of the method consists of multiplying the equation of state (EOS) by a factor smaller than one. This has been shown to improve the stability of lattice Boltzmann methods both in the context of pseudo potential methods by \citet{hu2013equations} as well as for free energy methods by Wagner and Pooley \cite{wagner2007interface}. This procedure leads to a wider interface thickness, better stability and smaller spurious currents magnitude. It is in some sense similar to reducing the time-step of the simulation, since the driving forces for pressure driven flows are similarly diminished. 

\citet{li2013lattice} performed simulations using the Carnahan-Starling (C-S) EOS and proposed a modification that is similar to multiply the EOS by a small factor (see Sec.~\ref{sec:Standard_Procedure} for more details). 
As a result, the authors obtained a wider interface thickness and a more stable method, which allowed the simulation of high-density ratios dynamic problems. This procedure has been widely applied in the LBM literature to improve the stability of simulations \cite{kharmiani2016simulation,fang2017lattice,pasieczynski2020multipseudopotential}.
Recently, \citet{peng2020attainment} obtained the most accurate planar interface results reported yet on the literature. 
They replaced the EOS van der Waals (vdW) loop (the EOS portion that consists in the meta-stable and unstable branches as shown in Fig.~(\ref{fig:VdWLoop}), a detailed description is provided in Sec.~\ref{sec:New_EOS})
by a cubic interpolation function and imposed the condition that disregarded the Maxwell equal area rule, and instead tuned it such the pseudopotential method recovers the desired vapor-liquid densities. This ensured excellent agreement between the target and measured coexistence plot. 
The key idea here is that the shape of the bulk pressure in the regions between the liquid and gas densities is irrelevant everywhere, except inside the smooth interface. These results again suggest that fine-tuning the van der Waals loop is a powerful tool to control the phase-coexistence and stability of a lattice Boltzmann method. 

The goal of this work is to further study how fine-tuning the van der Waals loop of an equation of state can be used to further improve the performance of pseudo potential lattice Boltzmann methods in terms of accuracy and stability. 
Following the work of \citet{peng2020attainment}, we use their strategy of replacing the EOS vdW loop by a smoother version. However, in contrast to the Peng method, we are ensuring that the sound speed changes continuously between the stable and meta-stable regions of the EOS and we ensure that this replacement will obey the Maxwell equal area rule.
This approach will be compared against the usual procedure of
multiplying the EOS by a factor smaller than one and also against the procedure proposed by Peng \textit{et al.} \cite{peng2020attainment}. 
Results show that the proposed vdW loop replacement is effective in increasing the numerical accuracy in static tests when compared with the unmodified vdW loop. 
%
%
Dynamic tests of a droplet impact in a liquid film with Reynolds number of 300 were performed to compare the stability of simulations carried with the C-S EOS, the C-S EOS with the Peng \textit{et al.} \cite{peng2020attainment} vdW loop replacement and also with the current proposed vdW loop replacement. Simulation parameters were selected to maintain the same interface width for all methods at each temperature. At these conditions the maximum stable density ratio achieved with 
the Peng \textit{et al.} \cite{peng2020attainment} vdW loop replacement
is 68. Under the same conditions, simulations with current proposed vdW replacement are able to remain stable for density ratios up to $3.38\cdot10^{4}$.

The current paper is organized as follows. 
In Sec.~\ref{sec:TheoreticalBackground}
the theoretical background related to LBM and the pseudopotential approach to multi-phase flows will be briefly discussed.
In Sec.~\ref{sec:PressureApproach} we compare static tests results using the C-S EOS (Secs.~\ref{sec:Regular_Pseudopotential} and \ref{sec:Standard_Procedure}), the C-S EOS with the Peng \textit{et al.} \cite{peng2020attainment} vdW loop replacement (Sec.~\ref{sec:Peng_EOS}) and also with the current proposed vdW loop replacement (Sec.~\ref{sec:New_EOS}). 
Then, in Sec.~\ref{sec:DynamicResults} dynamic simulations are carried to compare the performance of the different EOS strategies. Finally a brief conclusion drawn from numerical studies will be made in Sec.~\ref{sec:conclusion}.

\section{Theoretical Background}
\label{sec:TheoreticalBackground}


\subsection{The Lattice Boltzmann Equation} 
\label{sec:LatticeBoltzmannEquation}

The lattice Boltzmann equation (LBE) can be written as:

\begin{equation} 
\label{eq:LBE}
f_i(t+1,\bm{x}+\bm{c}_i) - f_i(t,\bm{x}) = \Omega_i(\bm{f},\bm{f}^{eq}) + F_i',
\end{equation}		
where $f_i$ are the particle distribution functions related with the velocity $\bm{c}_i$ and $f_i^{eq}$ are the local equilibrium distribution functions. The terms $\bm{f}$ and $\bm{f}^{eq}$ correspond to vectors whose components are $[\bm{f}]_i=f_i$ and $[\bm{f}]^{eq}_i=f^{eq}_i$.
Also, $t$ and $\bm{x}$ are the time and space coordinates, respectively. 
The term $\Omega_i(\bm{f},\bm{f}^{eq})$ is the collision operator and it is, in general, dependent on $\bm{f}$ and $\bm{f}^{eq}$. 
The multiple-relaxation time (MRT) collision operator is given by:
\begin{equation} 
\label{eq:OperatorMRT}
\Omega_i(\bm{f},\bm{f}^{eq}) = - \left[ \bm{M}^{-1} \bm{\Lambda} \bm{M} \right]_{ij}  (f_j - f_j^{eq}),
\end{equation}	
where $\bm{M}$ is the matrix that converts $(\bm{f} - \bm{f}^{eq})$ into a set of physical moments.
The velocity set used in this work is the regular two-dimensional nine velocities set (D2Q9):
\begin{equation} 
\label{eq:VelocitySet}
\bm{c}_i =
\begin{cases} 
      (0,0), ~~~~~~~~~~~~~~~~~~~~~~~~~~~~~~~~~~ i = 0,  \\
      (1,0), (0,1), (-1,0), (0,-1), ~~~~~ i = 1,...,4, \\
      (1,1), (-1,1), (-1,-1), (1,-1), ~ i = 5,...,8. \\
   \end{cases}
\end{equation}		
The specific form of $\bm{M}$ used in this work is \cite{d1992rarefied}:
\begin{equation} 
\label{eq:MatrixMRT}
\bm{M} = 
 	\begin{pmatrix}
 		1 & 1 & 1 & 1 & 1 & 1 & 1 & 1 & 1  \\
 		-4 & -1 & -1 & -1 & -1 & 2 & 2 & 2 & 2 \\ 
 		4 & -2 & -2 & -2 & -2 & 1 & 1 & 1 & 1 \\
 		0 & 1 & 0 & -1 & 0 & 1 & -1 & -1 & 1 \\
		0 & -2 & 0 & 2 & 0 & 1 & -1 & -1 & 1 \\
 		0 & 0 & 1 & 0 & -1 & 1 & 1 & -1 & -1 \\
 		0 & 0 & -2 & 0 & 2 & 1 & 1 & -1 & -1 \\
 		0 & 1 & -1 & 1 & -1 & 0 & 0 & 0 & 0 \\
 		0 & 0 & 0 & 0 & 0 & 1 & -1 & 1 & -1 \\
 		\end{pmatrix},
\end{equation}
the form of the matrix $M$ is choosen such that the collision matrix $\bm{\Lambda}$ responsible for the relaxation to local equilibrium becomes diagonal in the moment basis. $\bm{\Lambda}$ can be written as:
\begin{equation} 
\label{eq:RelaxationMatrix}
\bm{\Lambda} = \text{diag}\left( \tau_{\rho}^{-1},\tau_{e}^{-1},\tau_{\varsigma}^{-1},
\tau_{j}^{-1},\tau_{q}^{-1},\tau_{j}^{-1},\tau_{q}^{-1},
\tau_{\nu}^{-1},\tau_{\nu}^{-1} \right).
\end{equation}
where the parameters $\tau$ are the different relaxation times for the physical moments. The subscript of the relaxation times indicate the physical meaning of the moments. Of particular note are the conserved mass moment $\rho$ and the $x-$ and $y-$currents $j$, where the values of the relaxation times are arbitrary since the moments do not change in collisions. The moments related to trace of the stress tensor $e$ controlling the bulk viscosity as well as the remainder of the stress tensor $\nu$ controlling the shear viscosity. The three remaining moments, related to the symbols $\varsigma$ and $q$ are related to ``spurious moments" and are freely adjustable to improve the stability of the method.

The last term in the right-hand side of Eq.~(\ref{eq:LBE}), $F_i'$, 
is what defines the forcing scheme, 
i.e. this term is responsible for adding the effects of an external force field, $F_{\alpha}$, 
in the recovered macroscopic conservation equations. 
One of the most widely used forcing scheme in literature was developed by Guo \textit{et al.} \cite{guo2002discrete}.
The relation between particle distribution functions $f_i$ and the actual fluid velocity $\bm{u}$ depends on the forcing scheme.
For the Guo \textit{et al.} \cite{guo2002discrete} forcing scheme, density and velocity 
fields are given by:

\begin{subequations}
\begin{equation} 
\label{eq:Density}
\rho = \sum_i f_i,
\end{equation}	
\begin{equation} 
\label{eq:Momentum}
\rho \bm{u} = \sum_i f_i \bm{c}_i + \frac{\bm{F}}{2}.
\end{equation}	
\end{subequations}
The current $\bm{j}=\rho \bm{u}$ shown in Eq.~(\ref{eq:Momentum}) needs to take into account the force field
term, $\bm{F}/2$, as the average of the momentum before and after application of the forcing term. The Guo \textit{et al.} \cite{guo2002discrete} forcing scheme can be described as follows when the MRT collision operator is used: 

\begin{align}
F_i'  = &\left[ \bm{M}^{-1} \bigg( \bm{I} - \frac{\bm{\Lambda}}{2} \bigg) \bm{M} \right]_{ij}
w_j
\nonumber\\&
\times \left( \frac{c_{j \alpha}}{c_s^2} F_{\alpha}
+ \frac{(c_{j \alpha}c_{j \beta} - c_s^2 \delta_{\alpha \beta})}{c_s^4} 
F_{\alpha}u_{\beta} \right),
 \label{eq:GFS}
\end{align}	
where $\bm{I}$ is the identity matrix.

A popular form of the equilibrium distribution function $f_i^{eq}$ is \cite{kruger2017lattice}:
\begin{equation} 
\label{eq:EquilibriumDistribution}
f_i^{eq} = w_i \bigg( \rho + \frac{c_{i \alpha}}{c_s^2} \rho u_{\alpha}
+ \frac{(c_{i \alpha}c_{i \beta} - c_s^2 \delta_{\alpha \beta})}{2 c_s^4} 
\rho u_{\alpha}u_{\beta} \bigg),
\end{equation}	
where the terms $w_i$ are the weights related with each velocity $\bm{c}_i$, and $c_s$ is
the lattice sound speed. Note that this differs from the original definition by Qian \textit{et al.} \cite{qian1992lattice} in including the forcing correction of Eq. (\ref{eq:Momentum}).
The weights given by $w_i$
are $w_0 = 4/9$, $w_{1,2,3,4} = 1/9$ and $w_{5,6,7,8} = 1/36$. The sound speed $c_s$ can only take the value $1/\sqrt{3}$.

The LBE describes the evolution of particle distribution functions, however, the variables 
of interest are the macroscopic flow fields. The correspondence between the LBE and the
macroscopic behavior that it simulates can be shown through different approaches. 
The standard procedure is the Chapman-Enskog analysis, and one alternative is the 
recursive substitution developed by \citet{wagner1997theory} and further developed by 
\citet{holdych2004truncation} and \citet{kaehler2013derivation}. 
Up to second order terms, both procedures result in the same behavior, and it is not known if differences at higher orders will occur. 
Either approach recovers the mass and momentum conservation equations to second order:

\begin{subequations}
\begin{equation} 
\label{eq:MassConservation}
\partial_t \rho + \partial_{\alpha} (\rho u_{\alpha}) = 0,
\end{equation}	
\begin{eqnarray} 
\label{eq:MomentumConservation}
\partial_t (\rho u_{\alpha}) + \partial_{\beta} (\rho u_{\alpha} u_{\beta}) =
- \partial_{\alpha} (\rho c_s^2)
+ \partial_{\beta} \sigma_{\alpha\beta}' + F_{\alpha},
\end{eqnarray}	
\end{subequations}
where the viscous stress tensor, $\sigma_{\alpha\beta}'$, can be written as:
\begin{equation}
\label{eq:StressTensor}
    \sigma_{\alpha\beta}' = \mu (\partial_{\beta} u_{\alpha} + \partial_{\alpha} u_{\beta}) + \mu_B \delta_{\alpha \beta} \partial_{\gamma}u_{\gamma},
\end{equation}
and the dynamic viscosities $\mu$ and $\mu_B$ are related with the relaxation times of the LBM by:
\begin{equation}
\label{eq:DynamicViscosity}
    \mu = \rho c_s^2 \left( \tau_{\nu} - 0.5 \right) ~~~~~ 
    \mu_B = \rho c_s^2 \left( \tau_{e} - \tau_{\nu} \right),
\end{equation}

\subsection{Pseudopotential method} 
\label{sec:Pseudopotential}

The key idea of the pseudopotential method is that the external force in Eq. (\ref{eq:MomentumConservation}) can be used to model a non-ideal pressure by writing:
\begin{equation}
\label{eq:PressureForceRelation}
    - \partial_\beta p_{\alpha\beta} = - \partial_\alpha \rho c_s^2 +F_\alpha,
\end{equation}
an interaction force was proposed by \citet{shan1993lattice} based on nearest-neighbor interactions (see Shan \cite{shan2008pressure} for the definition of nearest-neighbor interactions):
\begin{equation}
\label{eq:ShanChenForce}
    F_\alpha^{SC} = - G \psi  ( \mathbf{x} )
\sum_i w ( | \mathbf{c}_i |^2 ) \psi ( \mathbf{x} + \mathbf{c}_i ) c_{i\alpha},
\end{equation}
where $\psi$ is a density-dependent interaction potential and $G$ is a parameter that controls the strength of interaction. The weights $w( | \bm{c}_i |^2 )$ are $w(1)=1/3$ and $w(2)=1/12$. The force, Eq.~(\ref{eq:ShanChenForce}), can be implemented into the LBE by using
different forcing schemes \cite{shan1993lattice,kupershtokh2009equations,he1998discrete}. The results shown in this section consider the use of forcing scheme proposed by the Guo \textit{et al.} \cite{guo2002discrete}.

According to \citet{shan2008pressure}, in the case of nearest-neighbor interactions, the pressure tensor (neglecting higher order terms) of the LBE resulting from the addition of the interaction force, Eq.~(\ref{eq:ShanChenForce}), is given by:
\begin{equation}
\label{eq:ShanPressureTensor}
    p_{\alpha\beta} = \left( 
    \rho c_s^2 + \frac{Gc^2}{2} \psi^2 + \frac{Gc^4}{12} \psi \partial_{\gamma} \partial_{\gamma} \psi 
    \right) \delta_{\alpha\beta} 
    + \frac{Gc^4}{6} \psi \partial_{\alpha} \partial_{\beta} \psi,
\end{equation}
where $c$ is the lattice constant. According to this result, the equation of state of the pseudopotential LB model is given by:
\begin{equation}
    p = \rho c_s^2 + \frac{Gc^2}{2} \psi^2,
\end{equation}
this expression motivated \citet{yuan2006equations} to use the effective density $\psi$ to add a new equation of state $p_{EOS}$ to the system:
\begin{equation}
\label{eq:InteractionPotential}
    \psi(\rho) = \sqrt{ \frac{2\left( p_{EOS} - \rho c_s^2 \right)}{G c^2} },
\end{equation}
when this technique is used, parameter $G$ no longer controls the interaction strength. If we replace Eq.~(\ref{eq:InteractionPotential}) into Eq.~(\ref{eq:ShanPressureTensor}) we can see that the dependence on $G$ is completely eliminated.
Now $G$ can be seen as an auxiliary parameter to keep the term inside the square root 
positive. Typically, the EOS parameters are set in such a way that $p_{EOS}<\rho c_s^2$. In this case, the value $G=-1$ can be adopted. 
For two phases in equilibrium separated by a planar interface, the Maxwell equal area rule states that the phase densities must satisfy the condition:
\begin{equation}
\label{eq:MaxwellRule}
\int_{v_l}^{v_v} \left( p_0 - p_{EOS} \right) d v =
\int_{\rho_v}^{\rho_l} \left( p_0 - p_{EOS} \right) \frac{d \rho}{\rho^2} = 0,
\end{equation}
%
where $\rho_v$ and $\rho_l$ are the saturated vapor and liquid densities. From the pressure tensor, Eq.~(\ref{eq:ShanPressureTensor}), an expression for the pseudopotential vapor-liquid relation can be derived \cite{shan2008pressure}:
\begin{align}
&\int_{\rho_g}^{\rho_l} \left( p_0 - \rho c_s^2 - \frac{Gc^2}{2} \psi^2 \right) 
\frac{\dot{\psi}}{\psi}d \rho
\nonumber\\
=& \int_{\rho_g}^{\rho_l} \left( p_0 - p_{EOS} \right) \frac{\dot{\psi}}{\psi}d \rho = 0,
\label{eq:ShanVaporLiqEq}
\end{align}
this expression is also called mechanical stability condition for the pseudopotential method.
As we can see this vapor-liquid density relation differs from the Maxwell rule Eq.~(\ref{eq:MaxwellRule}) unless a specific choice of $\psi$ is made, showing that the method is not thermodynamic consistent. 

The surface tension in a diffuse flat interface can be computed as the integral (along the normal direction in respect to the interface) of the mismatch between the normal $p_{xx}$ and transversal $p_{yy}$ components of the pressure tensor \cite{rowlinson2013molecular}. By computing the surface tension from Eq.~(\ref{eq:ShanPressureTensor}), we obtain:
\begin{equation}
\label{eq:IntegralSurfaceTension}
    \gamma = \int_{-\infty}^{+\infty} (p_{xx}-p_{yy}) dx = \frac{Gc^4}{6} \int_{-\infty}^{+\infty} (\psi \partial_{x} \partial_{x} \psi) dx,
\end{equation}
this result shows that the surface tension is not tunable independently of the equation of state ($\psi$ depends on $p_{EOS}$ as given by Eq.~(\ref{eq:InteractionPotential})) as stated by Sbragaglia \textit{et al.} \cite{sbragaglia2007generalized}.
If we had arbitrary coefficients in front of the gradient terms in the pressure tensor, Eq.~(\ref{eq:ShanPressureTensor}), instead of fixed coefficients, it would be possible to tune these parameters to change the surface tension given by the integral in Eq.~(\ref{eq:IntegralSurfaceTension}) independently on the choice of $\psi$. This procedure is similar to what is done in the Free energy literature, in the model developed by Swift \textit{et al.} \cite{swift1995lattice} the surface tension is adjusted by changing the parameter that multiply the gradient terms of the pressure tensor. This idea motivated Sbragaglia \textit{et al.} \cite{sbragaglia2007generalized} to propose a multi-range pseudopotential method that allows surface tension to be tuned.

With the introduction of multi-range interaction forces in \cite{shan2006analysis,sbragaglia2007generalized}, different forms of the pressure tensor than in the original method of  Eq.~(\ref{eq:ShanPressureTensor}) could be obtained. 
Also, different authors diverged about the form of the non-ideal pressure tensor that was resulting from the addition of an interaction force into the lattice Boltzmann method \cite{shan2008pressure,he2002thermodynamic}.
Although Eq.~(\ref{eq:PressureForceRelation}) is a physically consistent procedure in the macroscopic perspective to relate the non-ideal pressure with the interaction force, in the numerical viewpoint it does not consider the effect of discretization errors and the effect of higher order terms that not appear in the Chapman-Enskog expansion until second order.

\subsection{Recovering Thermodynamic consistency of the Shan-Chen model for arbitrary equations of state by altering the forcing term to obtain the correct pressure tensor}
\label{sec:RecoverPressureTensor}

Motivated by this issue Shan \cite{shan2008pressure} proposed a methodology to obtain the pressure tensor from a generic interaction force. The author also analyzed what is the form of the vapor-liquid relation for a more general pressure tensor than Eq.~(\ref{eq:ShanPressureTensor}), as is usually found in free energy approaches \cite{swift1995lattice}. Let's consider a pressure tensor with arbitrary coefficients $A$, $B$ and $C$:
\begin{align}
    p_{\alpha\beta} = \bigg( &
    \rho c_s^2 + \frac{Gc^2}{2} \psi^2 
    + A \frac{Gc^4}{12} ( \partial_{\gamma} \psi )( \partial_{\gamma} \psi )
    \nonumber\\&\left.
    + (B - \gamma) \frac{Gc^4}{12} \psi \partial_{\gamma} \partial_{\gamma} \psi 
    \right) \delta_{\alpha\beta} 
    + C \frac{Gc^4}{12} \psi \partial_{\alpha} \partial_{\beta} \psi,
\label{eq:GeneralPressureTensor}
\end{align}
following the work of Shan \cite{shan2008pressure}, it can be obtained a mechanical stability condition from this pressure tensor by imposing that $\partial_\alpha p_{\alpha\beta}=0$ in
equilibrium:
\begin{align}
&\int_{\rho_g}^{\rho_l} \left( p_0 - \rho c_s^2 - \frac{Gc^2}{2} \psi^2 \right) \frac{\dot{\psi}}{\psi^{1+\epsilon}}d \rho 
\nonumber\\
=& \int_{\rho_g}^{\rho_l} \left( p_0 - p_{EOS} \right) \frac{\dot{\psi}}{\psi^{1+\epsilon}}d \rho = 0,
\label{eq:VaporLiqEq}
\end{align}
where $\epsilon=-2A/B$, and $A$ and $B$ are the pressure tensor coefficients of Eq. (\ref{eq:GeneralPressureTensor}). In the case of the original Shan and Chen \cite{shan1993lattice} formulation, considering only nearest-neighbor interactions, the pressure tensor given by Eq.~(\ref{eq:ShanPressureTensor}) implies in $A=0$, $B=3$ and $\epsilon=0$, so Eq.~(\ref{eq:ShanVaporLiqEq}) is a particular case of Eq.~(\ref{eq:VaporLiqEq}). 
Based on the work of Shan \cite{shan2008pressure}, Li \textit{et al.} \cite{li2012forcing} proposed a modification in the Guo \textit{et al.} \cite{guo2002discrete} forcing scheme to introduce a free coefficient that could be used to change the vapor-liquid density relation and obtain coexistence densities close to the ones predicted by the Maxwell rule. Other forcing schemes were proposed in the literature \cite{li2013achieving,lycett2015improved,huang2016third} with the idea of obtain a pressure tensor in the form of Eq.~(\ref{eq:GeneralPressureTensor}) that allows the control of the coexistence curve and a surface tension that can be tunable independently on the equation of state. The same results can be obtained also by using multi-range interaction forces \cite{sbragaglia2007generalized}. In a recent work we found that a pressure tensor that allows the adjustment of the coexistence curve and independent surface tension control (without affect the interface width and coexistence densities) can be obtained by using only a nearest-neighbor interaction force without change the forcing scheme \cite{czelusniak2020force}.  

In this work we are going to use the forcing scheme proposed by \citet{li2013lattice} to control the coexistence curve and make the simulated densities satisfy the Maxwell rule. The reason for this choice is because this is the most used procedure in the literature and we want to show that shaping the EOS is a feature that can be used with any procedure to improve the numerical results in terms of accuracy and stability. 
The Li \textit{et al.} \cite{li2013lattice} forcing scheme is given by:
\begin{equation}
\label{eq:LiFS}
    \bm{\overline{F}}' = \bm{M}\bm{F}' = 
 	\begin{pmatrix}
 		0 \\
 		6 \left( u_x F_x + u_y F_y \right) 
 		+ \frac{12\sigma|\bm{F}|^2}{\psi^2 (\tau_e-0.5)}\\ 
 		- 6 \left( u_x F_x + u_y F_y \right)
 		- \frac{12\sigma|\bm{F}|^2}{\psi^2 (\tau_{\varsigma}-0.5)}\\
 		F_x \\
		-F_x \\
 		F_y \\
 		-F_y \\
 		2 \left( u_x F_x - u_y F_y \right) \\
 		u_x F_y + u_y F_x  \\
 		\end{pmatrix}.
\end{equation}
where $\sigma$ is a parameter used to tune $\epsilon$ in Eq.~(\ref{eq:VaporLiqEq}). According to \cite{li2013lattice} the new resulting pressure tensor have the following form:
\begin{equation}
\label{eq:NewPressureTensor}
    p_{\alpha\beta}^{new} = p_{\alpha\beta}^{Guo} 
    + 2 G^2 c^4 \sigma (\partial_{\gamma}\psi)(\partial_{\gamma}\psi) \delta_{\alpha\beta},
\end{equation}
where $p_{\alpha\beta}^{Guo}$ is the pressure tensor obtained with the Guo \textit{et al.} \cite{guo2002discrete} forcing scheme. When the Shan-Chen force is implemented with the Guo \textit{et al.} \cite{guo2002discrete} forcing scheme, the pressure tensor is given by Eq.~(\ref{eq:ShanPressureTensor}). In this case the $\epsilon$ parameter of Eq.~(\ref{eq:VaporLiqEq}), which is computed with the coefficients of the pressure tensor, is equal to $\epsilon=0$. If the Li \textit{et al.} \cite{li2013lattice} forcing scheme is used, the pressure tensor is given by Eq.~(\ref{eq:NewPressureTensor}) and the new $\epsilon$ parameter is $\epsilon=-16G\sigma$. Note that this relation is specific for the case when Shan-Chen force is used.
Other $\epsilon-\sigma$ relations relations could be obtained if different interaction forces were employed.

\subsection{Recovering correct liquid-gas coexistence values by changing the van-der-Waals loop of the equation of state}
A different strategy was proposed by \citet{peng2020attainment}. Starting from an arbitrary EOS, authors proposed the following modification:
\begin{equation}
\label{eq:PengEOS}
    \begin{aligned}
    p(\rho) & = p_{EOS}(\rho) ~~~~~~~~~~~~~~~~~~~~~~~~~~~~~~~~ \rho \leq \rho_{v}, \\
    p(\rho) & = p_0 + \theta(\rho-\rho_{v})(\rho-\rho_{l})(\rho-\rho_{m}) ~~~ \rho_{v} < \rho < \rho_{l}, \\
    p(\rho) & = p_{EOS}(\rho) ~~~~~~~~~~~~~~~~~~~~~~~~~~~~~~~~ \rho \geq \rho_{l},
    \end{aligned}
\end{equation}
where the parameter $\rho_{m}$ is computed numerically in order to satisfy the mechanical stability condition Eq.~(\ref{eq:VaporLiqEq}) for the densities given by applying the Maxwell rule to the original EOS. This EOS can then be applied directly using the Guo \textit{et al.} \cite{guo2002discrete} forcing scheme. 
The parameter $\theta$ can be used to control the interface width or surface tension (but not both independently).
In their work, Peng \textit{et al.} \cite{peng2020attainment} decided to related $\theta$ to the saturated vapor and liquid slopes by means of an auxiliary parameter $r_{\theta}$:
\begin{align}
    \theta = &\left. \frac{1-r_{\theta}}{(\rho_{v}-\rho_{m})(\rho_{v}-\rho_{l})} \frac{dp_{EOS}}{d\rho} \right|_{\rho = \rho_{v}}
    \nonumber\\&
    + \left. \frac{r_{\theta}}{(\rho_{l}-\rho_{m})(\rho_{l}-\rho_{v})} \frac{dp_{EOS}}{d\rho} \right|_{\rho = \rho_{l}}
\label{eq:PengTheta}
\end{align}

\section{Recovery of gas-liquid coexistence values for the different methods}
\label{sec:PressureApproach}

The accuracy of the pseudopotential method is a topic not well discussed in the literature.
When authors employ the Li \textit{et al.} \cite{li2013lattice} forcing scheme, they usually set the $\sigma$ parameter empirically to match the simulated phase densities with the ones given by the Maxwell rule \cite{li2012forcing,li2013lattice,kharmiani2016simulation,fang2017lattice}. Also, in some works authors matches phase densities of circular droplets with the densities given by the Maxwell rule. Since the Maxwell rule is only applicable to phases separated by a planar interface, it may not accurately predict phase densities in the case of curved interfaces \cite{czelusniak2020force}.
We judge that this procedure (of finding parameters empirically) can hide discrepancies between numerical results and what was intended in the model formulation. 
So, in this work the forcing scheme parameters will be calculated to make the theoretical mechanical stability condition, Eq.~(\ref{eq:VaporLiqEq}), be solved for the same densities that solves the Maxwell rule. In this way, we can see how numerical results are in agreement with the theoretically predicted coexistence curve and how higher order terms (which are difficult to be taken into account in a theoretical analysis) are influencing numerical results.

Following this approach of computing parameters from theory instead of empirically set them, we will first examine in Sec.~\ref{sec:Regular_Pseudopotential} the pseudopotential method results when the Guo \textit{et al.} \cite{li2013lattice} and Li \textit{et al.} \cite{li2013lattice} forcing schemes are employed. The observed deviations between the numerical densities and the desired thermodynamic consistent densities will motivate the study of how equations of state can be used to achieve a better agreement for the static coexistence densities. The standard procedure of change the EOS parameters to get a wider interface thickness will be analysed in Sec.~\ref{sec:Standard_Procedure}. Then, the method of change the EOS vdW loop proposed by Peng \textit{et al.} \cite{peng2020attainment} will be studied in Sec.~\ref{sec:Peng_EOS}. In the end of this section, a new EOS is proposed and results for a planar interface static test are showed in Sec.~\ref{sec:New_EOS}. As it will be shown in the dynamic texts (Sec.~\ref{sec:DynamicResults}) the motivation behind the proposition of a new EOS is the need to obtain a method that allow for a better agreement between the simulated densities with the Maxwell rule and able to remain stable in dynamic simulations under different conditions in terms of Reynolds number and density ratio.  

All simulations in this section were done considering the Carnahan-Starling (C-S) EOS which is given by: 
\begin{equation}
    \label{eq:CS}
    P_{C-S} = \rho R T \frac{1+b\rho/4+(b\rho/4)^2-(b\rho/4)^3}{(1-b\rho/4)^3}  - a \rho^2,
\end{equation}
with $b\approx0.5218/\rho_c$, $a\approx3.8532p_c/\rho_c$ and $R\approx2.7864p_c/(T_c\rho_c)$, where $\rho_c$, $p_c$ and $T_c$ are the critical density,
pressure and temperature, respectively. Thus, from the physical properties ($\rho_c$, $p_c$ and $T_c$) of the non-ideal gas (that we want to simulate) at the critical point we can define the parameter $a$, $b$ and $R$. Since in this work we are interested in study the numerical aspects of the method for a particular equation of state, we will simply pick values of $a$, $b$ and $R$ that are commonly used in the literature. 
In the pseudopotential literature
\cite{yuan2006equations,li2013lattice,kharmiani2016simulation}, authors 
commonly use the parameters $b=4$, $R=1$ and $a=1, 0.5$ or $0.25$ in numerical simulations.
In this study it will be adopted the values $a=0.5$, $b=4$ and $R=1$. 
The thermodynamic consistent densities are obtained from the EOS by numerically solving Eq.~(\ref{eq:MaxwellRule}).

\subsection{Regular Pseudopotential Method}
\label{sec:Regular_Pseudopotential}

In this subsection it will be shown how the pseudopotential method planar interface phase density values diverge from the expected ones 
for low values of reduced temperature. We will evaluate a system modelled by the Carnahan-Starling (C-S) EOS.
First we will implement the Shan-Chen \cite{shan1993lattice} force, Eq.~(\ref{eq:ShanChenForce}), using the Guo \textit{et al.} \cite{guo2002discrete} forcing scheme, Eq.~(\ref{eq:GFS}).
As discussed in Sec.~(\ref{sec:Pseudopotential}) this approach leads to a mechanical stability condition different from the 
Maxwell rule, unless the particular choice $\psi=\psi_0 exp(-\rho_0/\rho)$ is made. 
So, deviations of phase densities are expected for the C-S EOS.

Then we will evaluate how the Li \textit{et al.} \cite{li2013lattice} forcing scheme, Eq.~(\ref{eq:LiFS}), can be used to better approximate the coexistence curve given 
by the Maxwell rule. But in this work the $\sigma$ parameter (which is directly related to $\epsilon$ of Eq.~(\ref{eq:VaporLiqEq}) by $\epsilon=-16G\sigma$ for the Shan-Chen force)
will be set following a different procedure than it is done in many works in the literature \cite{li2012forcing,li2013lattice,kharmiani2016simulation,fang2017lattice}.
Instead of empirically set this parameter we will first calculate the $\epsilon$ values that solves the mechanical stability condition Eq.~(\ref{eq:VaporLiqEq}) (for each temperature) for the thermodynamic consistent densities. 
And then, the $\sigma$ parameter will be computed from these $\epsilon$ values.
In this way, we can show that numerical results (planar interface phase densities) given by the Li \textit{et al.} \cite{li2013lattice} forcing scheme also largely deviates from the behaviour predicted by the mechanical stability condition Eq.~(\ref{eq:VaporLiqEq}). 

The computational domain is given by a mesh of $(Nx,Ny)=(2,200)$ nodes.
The distribution function is initialized as equal to the equilibrium distribution function
$f_i(x,y,t=0)=f_i^{eq}[\rho(x,y),u_x\equiv 0,u_y\equiv 0]$ with null velocities. The density field is initialized as a diffuse planar interface given by the function: 
\begin{align}
\label{eq:PlanarIntDensity}
\rho(x,y) = \rho_{v} + \frac{\rho_l-\rho_v}{2} \bigg(& \text{tanh}  
\bigg[ \frac{4.6 (y-y_1)}{W} \bigg]
\nonumber\\&
- \left. \text{tanh}  
\bigg[ \frac{4.6 (y-y_2)}{W} \bigg] \right),
\end{align}
where $y_1=0.25Ny$ and $y_2=0.75Ny$ are the location of the interfaces. 
Also, $W$ is the interface width. In this work we consider the interface as the region where the density has values inside the range $0.01 < (\rho-\rho_v)/(\rho_l-\rho_v) < 0.99$. This definition is consistent with the choice of the factor 4.6 in Eq.~(\ref{eq:PlanarIntDensity}).
For the planar interface initialization we set $W=10$. This value is adopted
only for initialization purposes, for each simulation a specific equilibrium width can be obtained depending on the physical conditions such as the reduced temperature.
For a specific temperature, the values of $\rho_v$ and $\rho_l$
are initialized as the saturated densities obtained with the Maxwell equal area rule. 
We are using periodic boundary conditions for the distribution function. 

The LBE was solved using the MRT collision operator and all relaxation times in Eq.~(\ref{eq:RelaxationMatrix}) are made equal to one. 
Setting all the relaxation times with the same value reduces the MRT to be equal to the BGK collision operator. Different relaxation times will be used in the dynamic tests, where only $\tau_{\nu}$ will be adjusted to control the shear viscosity without change the values of other relaxation times which we expect to result in better stability than standard BGK. 
Simulations are carried until the following convergence criteria has being obeyed:
\begin{equation} 
\frac{ \sum_x \mid [\rho(x,t)-\rho(x,t-100)] \mid }{ \sum_x \mid \rho(x,t) \mid } < 10^{-6}.
\end{equation}
As discussed in Sec.~(\ref{sec:Pseudopotential}) we can set $G=-1$ in Eq.~(\ref{eq:InteractionPotential}).

\begin{figure}
\centering
	\includegraphics[width=80mm]{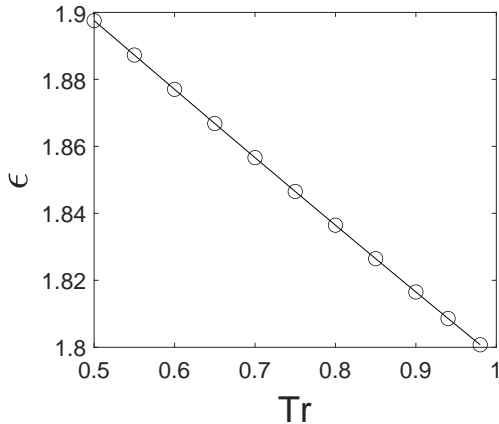}
	\caption{Curve of $\epsilon$ values numerically obtained from solving Eq.~(\ref{eq:VaporLiqEq}) for the C-S EOS with $a = 0.5$ at different reduced temperatures.}
	\label{fig:e_CS}
\end{figure}

Now, we need to show how the $\sigma$ parameter of the Li \textit{et al.} \cite{li2013lattice} forcing scheme is specified. Remember that $\sigma$ is a parameter of the forcing scheme, Eq.~(\ref{eq:LiFS}), and by changing it we modify the resulting pressure tensor. Since $\epsilon$ in Eq.~(\ref{eq:VaporLiqEq}), which is the parameter that controls the coexistence curve, is computed with the pressure tensor coefficients, we can use $\sigma$ to change the pressure tensor and adjust $\epsilon$.
In many works, this $\sigma$ parameter is set empirically, but in this work it will be calculated.
For each temperature, the Maxwell densities are used in Eq.~(\ref{eq:VaporLiqEq}) and starting from an initial guess $\epsilon=2$, the correct value of this parameter that solves the vapor-liquid relation is obtained by the Newton method which is an iterative procedure. The Simpson's rule \cite{atkinson2008introduction} is used to perform the numerical integration. The computed values of $\epsilon$ are shown in Fig.~(\ref{fig:e_CS}), for this range of reduced temperatures the curve of $\epsilon$ parameters for the C-S EOS approaches a straight line. The reduced temperature is commonly defined as $T_r=T/T_c$.
Then, we compute $\sigma$ using $\epsilon=-16G\sigma$, as discussed in Sec.~\ref{sec:RecoverPressureTensor}.

\begin{figure}
	\centering
	\includegraphics[width=\columnwidth]{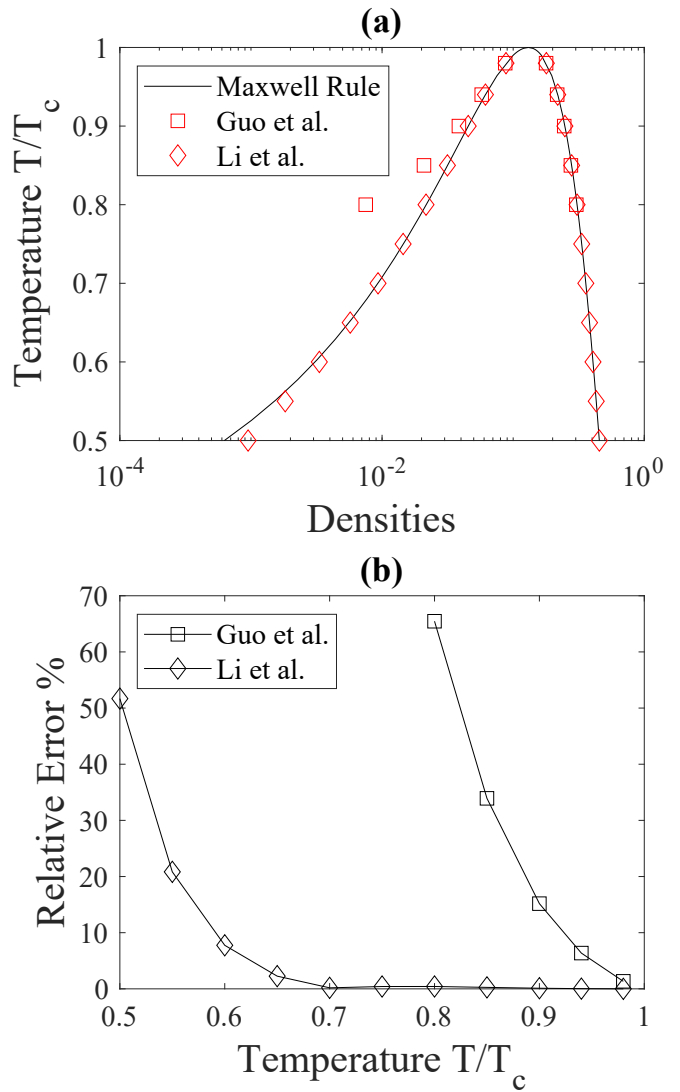}
	\caption{
	(a) Coexistence curve for the C-S EOS (using $a=0.5$) with the Shan-Chen force, Eq.~(\ref{eq:ShanChenForce}) implemented using the Guo \textit{et al.} 
	\cite{guo2002discrete} and Li \textit{et al.} \cite{li2013lattice} forcing scheme.
	(b) Relative error for the numerical vapor densities in respect to the coexistence vapor densities predicted by the Maxwell rule, Eq.~(\ref{eq:DensityError}). 
	}
	\label{fig:Theory_CS}
\end{figure}

We constructed two numerical coexistence curves. In one of them, the Shan-Chen force is implemented with the forcing scheme given by Eq.~(\ref{eq:GFS}) and the other with 
the forcing scheme defined by Eq.~(\ref{eq:LiFS}). The results are shown 
in Fig.~(\ref{fig:Theory_CS}.a). The numerical coexistence curves are also compared with the coexistence curve given by the Maxwell rule applied to the C-S EOS. 
In Fig.~(\ref{fig:Theory_CS}.b)
we show the relative error of the simulated vapor density $\rho_v$ in comparison with the saturated vapor density given by the Maxwell rule $\rho_v^{M}$. The error is computed using the following expression:
\begin{equation}
\label{eq:DensityError}
    error = 100(\rho_v-\rho_v^{M})/\rho_v^{M}
\end{equation}
We decided to compute the error in terms of the vapor density, since errors related with
the liquid density are expected to be much smaller. Due to the mechanical equilibrium, the saturated vapor and liquid pressure must be equal. Since the liquid phase is much more incompressible than the vapor phase, a deviation in this saturation pressure represents a much higher deviation in the vapor density than in the liquid phase. For instance, lets consider the C-S EOS and a reduced temperature $T_r=0.5$. The saturation pressure at this temperature is $P_{sat}\approx1.5\cdot10^{-5}$. If we consider a small deviation of $+0.1$ \% in the liquid density the new saturation pressure increases to $P_{sat}\approx2\cdot10^{-4}$ what would require a change of $+1500$ \% in the vapor density. 

When the Guo \textit{et al.} \cite{guo2002discrete} forcing scheme is used, the vapor density values greatly diverge from the ones given by the Maxwell rule as seen in Fig.~(\ref{fig:Theory_CS}). 
Simulations also became unstable for reduced temperatures smaller than 0.8. A much better agreement is obtained when the Li \textit{et al.} \cite{li2013lattice} forcing scheme is used, the relative errors are very small for reduced temperatures above $T_r=0.7$. 
But below this temperature the error start to rapidly increase and reach approximately $50$\% when $T_r=0.5$.

The vapor density deviations in respect to the Maxwell rule for the Guo \textit{et al.} \cite{guo2002discrete} forcing scheme were expected since the mechanical stability condition
Eq.~(\ref{eq:ShanVaporLiqEq}) is different from Eq.~(\ref{eq:MaxwellRule}) 
in the case where the C-S EOS is chosen. 
But for the Li \textit{et al.} \cite{li2013lattice} forcing scheme, the mechanical stability condition was adjusted to match with the results given by the Maxwell rule. 

This solved the issue for reduced temperatures above 0.7, but errors were still observed for lower temperatures. These results indicate that higher order terms that are difficult to be taken into account in theoretical analysis are playing an important role in these lower temperatures and deviates the method physical behaviour from what it was expected. In the rest of this section it will be studied how equations of state can be used to overcome this issue. 

\subsection{Increasing interface thickness to improve numerical accuracy}
\label{sec:Standard_Procedure}

It was originally observed for free energy approaches to multiphase lattice Boltzmann methods that multiplying the EOS by a factor smaller than one can increase stability and affects the interface witdth and surface tenstion as well as the sound speed \cite{wagner2007interface}. 
Hu \textit{et al.} \cite{hu2013equations} showed that this approach carries over to pseudopotential methods where multiplying the equation of state 
by a small factor similarly increases the achievable density ratio, 
reduces spurious currents and controls surface tension. 
It was also observed that this procedure can be used to change the interface width \cite{lycett2015improved}, but not independently of the surface tension. 
One approach to allow independent control of surface tension is adopt an appropriate forcing scheme \cite{li2013achieving,lycett2015improved} or an interaction force \cite{czelusniak2020force}. 
An alternative approach to multiply the EOS by a small factor and that is often used in the pseudopotential literature was proposed by Li \textit{et al.} \cite{li2013lattice} and consists in reduce the $a$ parameter in the equation of state Eq.~(\ref{eq:CS}). The authors showed that this procedure is equivalent to multiply the EOS by a small factor. We can easily check this by writing Eq.~(\ref{eq:CS}) in terms
of the reduced temperature $T=T_rT_c$. Since $T_c \propto a$ (we can write $T_c\approx0.3773a/(Rb)$), we can isolate $a$ in front of the equation. 
In this way, by changing the $a$ parameter we are modifying the equation of state, but without affecting the coexistence densities given by the Maxwell rule at the same reduced temperatures.  

We are going to apply different modifications to the EOS, so for simplicity we will call the strategy of multiplying the EOS by a small factor, or equivalently reducing the $a$ parameter, the \textit{standard procedure}. In the previous subsection we showed that applying the Guo \textit{et al.} \cite{guo2002discrete} forcing scheme to the pseudopotential force fails to reproduce the Maxwell rule. We also showed
that the Li \textit{et al.} \cite{li2013lattice} forcing scheme
provided better agreement with the Maxwell rule, but still showed discrepancies in the phase densities for low reduced temperatures. This discrepancy can be attributed to the effect of higher order terms which are difficult 
to take into account.
In this section we are going to evaluate how the standard procedure affects the numerical coexistence curve and if it can approach the simulated results with the ones predicted by the mechanical stability condition. 

We selected three values of "a" and performed the planar interface tests using the same procedure applied in Sec.~\ref{sec:Regular_Pseudopotential} for the Li \textit{et al.} \cite{li2013lattice} forcing scheme. The "a" parameters were selected in such a way that for a temperature of $T_r=0.5$ the diffuse planar interface have the measured interface widths (measured when simulation reaches the equilibrium) of $w=7$, $9$ and $11$ as shown in Table~(\ref{tab:AparamStandard}). We are using the same definition of interface width used in Sec.~\ref{sec:Regular_Pseudopotential} which is the diffuse interface portion where $\rho$ is inside the range $0.01 < (\rho-\rho_v)/(\rho_l-\rho_v) < 0.99$. But here, $\rho_v$ and $\rho_l$ are the coexistence densities given by the simulated results which can be different from the ones given by the Maxwell rule. 
\begin{table}
\centering
\caption{\label{tab:AparamStandard}Values of the C-S EOS $a$ parameters selected to achieve the desired interface widths $w=7$, $9$ and $11$ for a reduced temperature of $T_r=0.5$. 
}
\begin{ruledtabular}
\begin{tabular}{cccc} 
&
\textrm{$w=7$}&
\textrm{$w=9$}&
\textrm{$w=11$}\\
\hline
$a$ & 0.363 & 0.215 & 0.141
\end{tabular}
\end{ruledtabular}
\end{table}

\begin{figure}
\centering
	\includegraphics[width=\columnwidth]{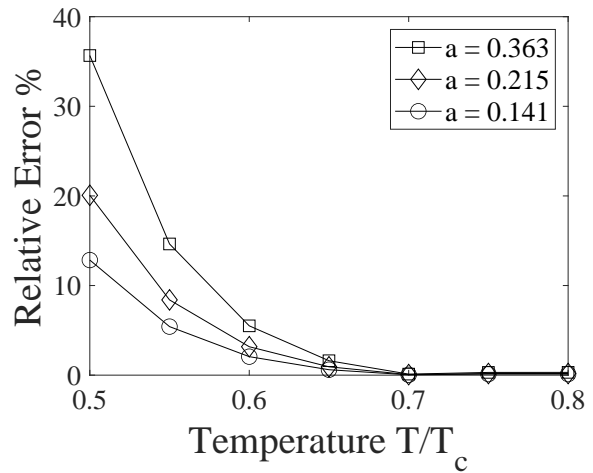}
	\caption{
	Relative error for the simulated vapor densities, obtained using the C-S EOS with different $a$ parameters, in respect to the saturated vapor densities predicted by the Maxwell rule, Eq.~(\ref{eq:DensityError})
	}
	\label{fig:Saturation_Numerical_CS}
\end{figure}

The temperature $T_r=0.5$ was used as a reference to measure the interface width, but the same values of $a$ will be used for simulations at other temperatures.
The planar interface results for this case are shown in Fig.~(\ref{fig:Saturation_Numerical_CS}). The vapor density relative error decreases when the interface width is increased. For the reduced temperature $T_r=0.5$ when $a=0.363$ and $w=7$, the relative vapor density error was close to $35$\%. This error reduced to about $12$\% for $a=0.141$ and $w=11$. 

\subsection{Peng EOS}
\label{sec:Peng_EOS}

As discussed in Sec.~\ref{sec:Regular_Pseudopotential}, the 
Li \textit{et al.} \cite{li2013lattice} forcing scheme requires the user to set the 
$\sigma$ parameter which have impact on the coexistence density values. 
Usually in the literature, authors adjust this parameter empirically until the solution
of a planar interface or static droplet test matches with the desired densities. 
In order to avoid this trial and error procedure, Peng \textit{et al.} \cite{peng2020attainment}
developed an alternative procedure to adjust the pseudopotential coexistence curve without introduce
modifications in the forcing scheme or in the interaction force. 
Authors proposed to change the EOS van der Waals loop, 
to a new one given by Eq.~(\ref{eq:PengEOS}). 
In this method, the Guo \textit{et al.} \cite{li2013lattice} forcing scheme it is used.
The bulk regions stay intact and the $\rho_m$ parameter in the vdW loop can be defined to make the mechanical stability condition, Eq.~(\ref{eq:ShanVaporLiqEq}), be satisfied for the thermodynamic consistent densities. In this way, parameter $\rho_m$ can be computed numerically avoiding any empirical tuning. 

In this work, we showed that is not necessary to find the $\sigma$ parameter empirically, it is possible to obtain it directly from the mechanical stability condition, Eq.~(\ref{eq:VaporLiqEq}), by relating $\sigma$ with $\epsilon$. But this procedure works well for reduced temperatures above to $T_r=0.7$ as observed in Fig.~(\ref{fig:Theory_CS}). For lower temperatures, the simulated results diverge from the mechanical stability condition. In the other hand, Peng \textit{et al.} \cite{peng2020attainment} observed that their method provided excellent agreement between numerical results and the Maxwell rule. 

In this subsection we test the Peng \textit{et al.} \cite{peng2020attainment} method and compare with the results obtained for the standard procedure. The vdW loop of the C-S EOS with $a=0.5$ (other parameters are equal to the ones used in Sec.~\ref{sec:Regular_Pseudopotential}) is replaced by Eq.~(\ref{eq:PengEOS}). 
Three parameters $r_{\theta}$ are selected. They were chosen in such a way that for a temperature of 
$T_r=0.5$ the diffuse planar interface have the measured interface widths (measured when simulation reaches the equilibrium) of $w=7$, $w=9$ and $11$ as shown in Table~(\ref{tab:ParamPeng}).
\begin{table}[htbp]
\centering
\caption{\label{tab:ParamPeng} Values of the Peng C-S EOS $r_{\theta}$ parameters selected to achieve the desired interface widths $w=7$, $9$ and $11$ for a reduced temperature of $T_r=0.5$. 
}
\begin{ruledtabular}
\begin{tabular}{cccc}
&
\textrm{$w=7$}&
\textrm{$w=9$}&
\textrm{$w=11$}\\
\hline
$r_{\theta}$ & 0.290 & 0.122 & 0.040
\end{tabular}
\end{ruledtabular}
\end{table}

\begin{figure}
\centering
	\includegraphics[width=\columnwidth]{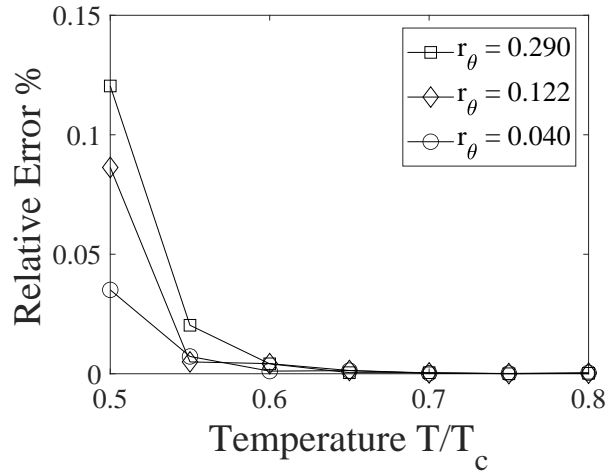}
	\caption{
	Relative error for the vapor densities at different reduced temperatures obtained using the Peng \textit{et al.} \cite{peng2020attainment} replacement for the vdW loop with different $r_{\theta}$.
	}
	\label{fig:Saturation_Numerical_Peng_CS}
\end{figure}

The temperature $T_r=0.5$ was used as a reference to measure the interface width, but the same values of $r_{\theta}$ will be used for simulations at other temperatures.
The relative errors of the vapor density, Eq.~(\ref{eq:DensityError}), for planar interface simulations using Peng EOS with different $r_{\theta}$ are shown in 
Fig.~(\ref{fig:Saturation_Numerical_Peng_CS}).
For a reduced temperature of $T_r=0.5$ and $r_{\theta}=0.290$ which gives an interface width of $w=7$, the error was close to $0.12$\% while for higher interface widths the error was smaller then $0.1$\%.
The errors obtained using the Peng EOS with the 
Guo \textit{et al.} \cite{li2013lattice} forcing scheme were significantly smaller then 
the ones obtained with the standard approach.

\subsection{New EOS}
\label{sec:New_EOS}

In this work a third strategy will be devised. It will be shown later
that although the Peng method greatly improved the static tests results, some issues concerning 
the method stability were observed in the dynamic tests. In this way, it is necessary to find a strategy
that combines better accuracy in respect with the coexistence curve and also capable of handle with dynamic simulations under high values of Reynolds number and large density ratios.  
In the new EOS,
instead of multiplying the entire equation by a small factor, only the van der Waals loop will be replaced by a smoother function following the work of Peng \cite{peng2020attainment}. But now, differently from the 
Peng \cite{peng2020attainment} work,
the EOS must satisfy the Maxwell rule and in this way the new EOS should be implemented using the Li \textit{et al.} \cite{li2013lattice} forcing scheme. Another condition that will be imposed to the EOS is that not only the pressure must be continuous but also the sound speed have to be continuous. 

\begin{figure}
\centering
	\includegraphics[width=70mm]{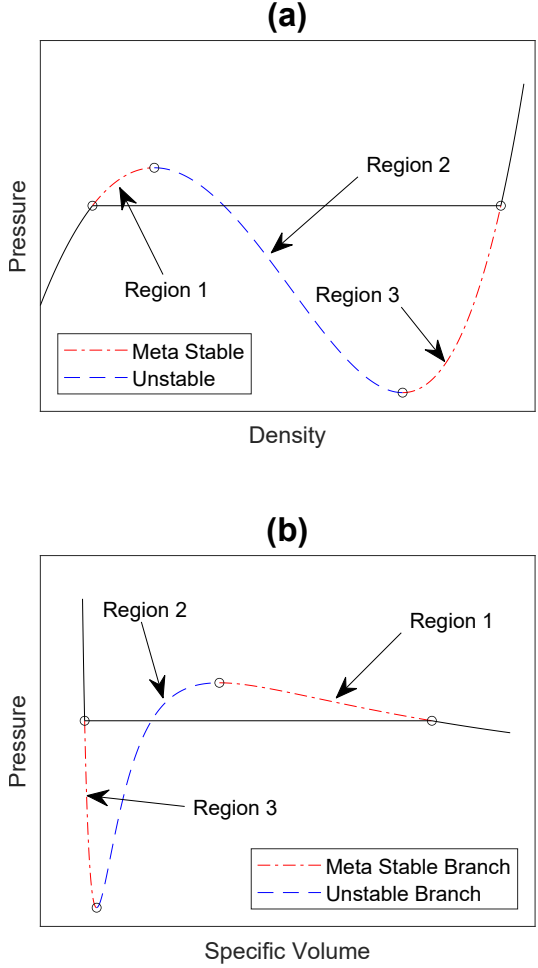}
	\caption{Illustration of the Van der Waal's loop for a generic EOS at a specific $T_r$. (a) Pressure-density relation $P_{EOS}=P(\rho,T_r)$. (b) Pressure-volume relation $P_{EOS}=P(v,T_r)$.}
	\label{fig:VdWLoop}
\end{figure}

Before describing how the new equation of state is devised it is worth it to define the van der Waals loop.
The form of a generic non-ideal gas pressure-density and pressure-volume relation for a certain reduced temperature is shown in Figs.~(\ref{fig:VdWLoop}.a) and (\ref{fig:VdWLoop}.b). The saturation pressure (and also the vapor and liquid densities) is given by the Mawell equal area rule. The unstable branch is the EOS portion where $\partial P/\partial v >0$ and the meta stable branches are the regions in between the saturation condition and the unstable branches. The van der Waals loop consist in the region of the EOS composed by the meta stable and unstable branches. Now, the new EOS will be described in details in the rest of this subsection.

In Fig.~(\ref{fig:VdWLoop}.a), the vdW loop was divided in three regions included within four points. 
The first region consist in the meta stable branch between the point defined by the vapor density and pressure ($\rho_{v},p_{v}=p_{sat}$) and ends in the point where the pressure is maximum ($\rho_{max},p_{max}$) inside the vdW loop. The second region is the unstable branch and goes from the point of maximum ($\rho_{max},p_{max}$) to the point of minimum pressure ($\rho_{min},p_{min}$) inside the vdW loop where starts the third region that ends in the point that defines the liquid phase ($\rho_{l},p_{l}=p_{sat}$). Now, it is possible to create a customized vdW loop by replacing these three regions by interpolating functions passing through these four points. 

The region 1 will be replaced by an elliptic interpolating function:
\begin{subequations}
\begin{equation}
    p(\rho) = p_{max} - b_1 + b_1 \sqrt{ 1 - \frac{(\rho - \rho_{max})^2}{a_1^2} };
    ~~~
    \rho_v < \rho \leq \rho_{max},
\end{equation}
\begin{equation}
    a_1^2 = - \frac{k^2}{2k + (\rho_{max}-\rho_{v})^2},
\end{equation}
\begin{equation}
    k = \frac{(p_{max}-p_{sat})(\rho_{max}-\rho_{v})}
    {\frac{\partial p}{\partial \rho} (\rho_{v})} - (\rho_{max}-\rho_{v})^2,
\end{equation}
\begin{equation}
    b_1 = \frac{p_{max}-p_{sat}}{1-\sqrt{1-\frac{(\rho_{max}-\rho_{v})^2}{a_1^2}}}.
\end{equation}
\end{subequations}

\begin{figure}[H]
\centering
	\includegraphics[width=\columnwidth]{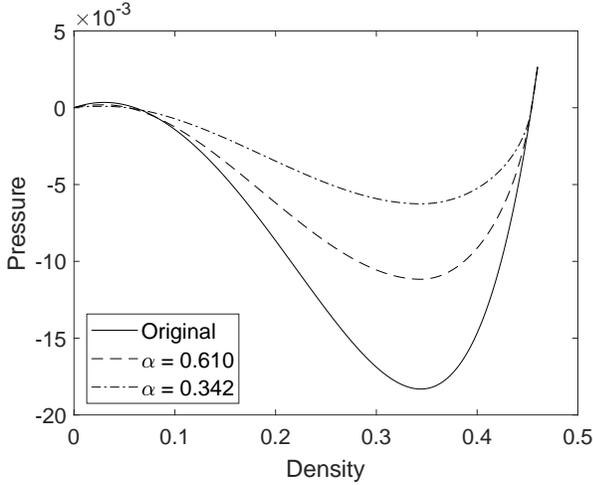}
	\caption{Pressure-density relation for the Smooth C-S.}
	\label{fig:Smooth_EOS}
\end{figure}

In order to $a_1$ be a real number, the following condition must be satisfied: $\partial p/\partial \rho (\rho_{v}) > 2(p_{max}-p_{sat})/(\rho_{max}-\rho_{v})$. In this work we adopt $\rho_{max}$ as equal to the density where the pressure is maximum
in the original C-S EOS $\rho_{max}=\rho_{max}^{C-S}$, for the same temperature. We observed that this option did not violated the condition
for $a_1$ be a real number. 
The region 2, which is the unstable branch, will be replaced by a cubic polynomial interpolating function:
\begin{subequations}
\begin{equation}
\begin{aligned}
    p(\rho) &= a_2 (\rho-\rho_{max})^3 + b_2 (\rho-\rho_{max})^2 \\ 
    &+ c_2 (\rho-\rho_{max}) + d_2;
    ~~~
    \rho_{max} < \rho < \rho_{min},
\end{aligned}
\end{equation}
\begin{equation}
    \begin{aligned}
    d_2 = & ~ p_{max}, \\
    c_2 = & ~ 0, \\
    b_2 = & ~ 3(p_{min} - p_{max}), \\
    a_2 = & ~ 2(p_{max} - p_{min}),
    \end{aligned}
\end{equation}
\end{subequations}
where $\rho_{min}$ is adopted as equal to the density where the pressure is minimum
in the original C-S EOS, $\rho_{min}=\rho_{min}^{C-S}$, for the same temperature.
The region 3 will be also replaced by an elliptic interpolating function:
\begin{subequations}
\begin{equation}
    p(\rho) = p_2 + b_3 - b_3 \sqrt{ 1 - \frac{(\rho - \rho_2)^2}{a_3^2} };
    ~~~
    \rho_{min} \leq \rho < \rho_l,
\end{equation}
\begin{equation}
    a_3^2 = - \frac{k^2}{2k + (\rho_{l}-\rho_{min})^2},
\end{equation}
\begin{equation}
    k = \frac{(p_{sat}-p_{min})(\rho_{l}-\rho_{min})}
    {\frac{\partial p}{\partial \rho} (\rho_{l})} - (\rho_{l}-\rho_{min})^2,
\end{equation}
\begin{equation}
    b_3 = \frac{p_{sat}-p_{min}}{1-\sqrt{1-\frac{(\rho_{l}-\rho_{min})^2}{a_3^2}}}.
\end{equation}
\end{subequations}
The following parameters are imposed to be equal to the ones from the original EOS:
\begin{equation}
    \rho_{v}; ~~ \rho_{min}; ~~ \rho_{max}; ~~ \rho_{l}; 
    ~~
    p_{v} = p_{l} = p_{sat};
    ~~
    \frac{\partial p}{\partial \rho} (\rho_{v});
    ~~
    \frac{\partial p}{\partial \rho} (\rho_{l}),
\end{equation}
the minimum pressure $p_{min}$ will be a free parameter that can be chosen in order to control the degree of smoothness of the vdW loop. This pressure will be related to the original C-S minimum pressure using a parameter $\alpha$ as $p_{min}=p_{sat}-\alpha(p_{sat}^{C-S}-p_{min}^{C-S})$. The only parameter that was not specified yet is the maximum pressure $p_{max}$ which is computed using the requirement that the new equation of state must satisfy the Maxwell rule, Eq.~(\ref{eq:MaxwellRule}), for $\rho_v$ and $\rho_l$ given by the original C-S EOS.
The value of $p_{max}$ is obtained numerically. 
This new EOS that is a modification of the original C-S EOS will be called along the text as "Smooth C-S".
As an example, from the original C-S EOS, with $a=0.5$, it will be derived two modified versions, one with $\alpha=0.610$ and other with $\alpha=0.342$. In Fig.~(\ref{fig:Smooth_EOS}) it is possible to see how the Smooth C-S looks like for these $\alpha$ parameters. 

Now we are going to evaluate how the Smooth C-S EOS affects the numerical coexistence curve for the pseudopotential LBM. We selected three values of $\alpha$ and again this parameters were selected in such a way that for a temperature of $T_r=0.5$ the diffuse planar interface have the measured interface widths 
of $w=7$, $9$ and $11$ as shown in Table~(\ref{tab:ParamSmooth}). 
\begin{table}[htbp]
\centering
\caption{\label{tab:ParamSmooth}Values of the Smooth C-S EOS $\alpha$ parameters selected to achieve the desired interface widths $w=7$, $9$ and $11$ for a reduced temperature of $T_r=0.5$. 
}
\begin{ruledtabular}
\begin{tabular}{cccc}
\textrm{$\phantom{L}$}&
\textrm{$w=7$}&
\textrm{$w=9$}&
\textrm{$w=11$}\\
\hline
$\alpha$ & 0.610 & 0.342 & 0.216
\end{tabular}
\end{ruledtabular}
\end{table}

\begin{figure}
\centering
	\includegraphics[width=\columnwidth]{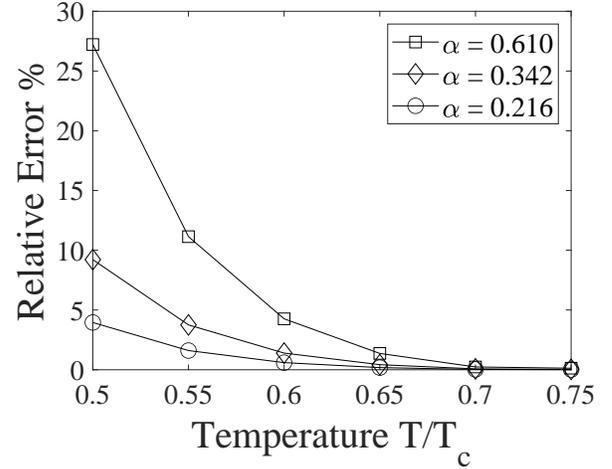}
	\caption{
	Relative error for the vapor densities at different reduced temperatures obtained using the Smooth C-S EOS with different $\alpha$ parameters.
	}
	\label{fig:Saturation_Numerical_Smooth}
\end{figure}

The temperature $T_r=0.5$ was used as a reference to measure the interface width, but the same values of $\alpha$ will be used for simulations at other temperatures.
Results for the Smooth EOS are shown in Fig.~(\ref{fig:Saturation_Numerical_Smooth}). We can see that
errors are smaller compared with original C-S EOS when both EOS
are applied using the same forcing scheme \cite{li2013lattice}.
When we reduce the $a$ parameter of the C-S EOS, the effect is the same as multiply it by a small factor. Which means that the vapor and liquid region sound speeds are also reduced. In the smooth EOS, the bulk regions are unchanged, and only the van der Waals loop is replaced. So, the sound speeds of the vapor and liquid regions are higher compared with the original EOS multiplied by a small factor. 

\begin{figure}
\centering
	\includegraphics[width=\columnwidth]{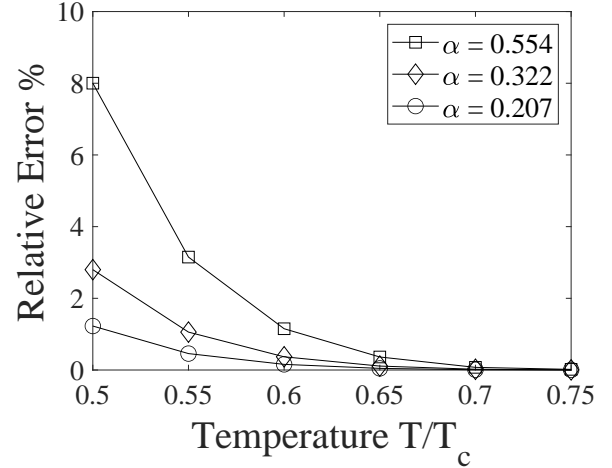}
	\caption{
	Relative error for the vapor densities at different reduced temperatures obtained using the Smooth C-S EOS (with higher vapor sound speed) with different $\alpha$ parameters.
	}
	\label{fig:Saturation_Numerical_Smooth_high_slope}
\end{figure}

In order to test the influence of the vapor sound speed in the density errors we create a Smooth EOS where the vapor region is given by the C-S EOS with $a=2$. The liquid pressure is still given by the C-S EOS with $a=0.5$, but a term $\Delta P$ is added to account to the new saturation pressure. In this way we changed the vapor region sound speed without change the liquid sound speed. 
Results are shown in Fig.~(\ref{fig:Saturation_Numerical_Smooth_high_slope}).
We conclude that by increasing the vapor region sound speed, the density errors are further reduced. The new parameters $\alpha = 0.554, 0.322$ and $0.207$ were computed in other to maintain the interface width as $w=7, 9$ and $11$, respectively, when $T_r=0.5$. 
These $\alpha$ parameters were computed in respect to the minimum pressure of the C-S EOS with $a=0.5$.
The static test results in Fig.~(\ref{fig:Saturation_Numerical_Smooth_high_slope}) showed that the we can increase the vapor region sound speed, which also makes the vapor region more incompressible, to increase the method accuracy at lower reduced temperatures.

\section{Numerical results of Dynamic Tests}
\label{sec:DynamicResults}

In this section, the LBM performance using different EOS will be evaluated in the dynamic test of a droplet splashing on a thin liquid film. 
We developed a
finite difference (FD) method that replicates the solution of the pseudopotential lattice 
Boltzmann method. With this FD method it is possible to perform a mesh refinement and
obtain reference solutions for the simulations done in this section. Details about how this FD method was devised are shown in Appendix \ref{sec:AppendixFDmethod}.

The physical domain is given by a rectangle of $(L_x,L_y)=(600,250)$. If the LBM is used $\Delta x=1$, for the FD method it will be used $\Delta x = 1/2$ as a reference. A study about the mesh impact on the FDM results is shown in Appendix \ref{sec:GridStudy}. Periodic boundary conditions are applied in the left and right side and non-slip condition is used in the top and bottom walls. For the non-slip boundary conditions it is used the bounce-back procedure. In respect with the fluid-solid interaction, for simplicity at the walls the interaction force acting on the fluid is computed only using Eq.~(\ref{eq:ShanChenForce}), but considering that the nodes inside the solid wall have a density equal to the saturated liquid (bottom wall) or vapor (top wall). Above the bottom wall there is a liquid film with high $h_y=25$. A droplet of radius $R=50$ and velocity $(v_x,v_y)=(0,-V_d)$ is initialized above the liquid film. The droplet and the liquid film are initialized using the hyperbolic tangential function as it was done for the planar interface test.
The interface width is initialized equal to the one measured in the planar interface test for the same reduced temperature.
The LBE is solved using the MRT collision operator with the relaxation matrix:
\begin{equation} 
    \bm{\Lambda} = 
    \text{diag} \left( 1,1,1,1,1,1,1,\tau^{-1},\tau^{-1} \right),
\end{equation}
where $\tau$ is dependent on the density ($\tau_v$ and $\tau_l$ are the relaxation times for the vapor and liquid phase):
\begin{equation}
\label{eq:RelaxationTime}
    \tau = \tau_l \frac{\rho - \rho_v}{\rho_l - \rho_v}
            + \tau_v \frac{\rho - \rho_l}{\rho_v - \rho_l}
\end{equation}
also, $\tau_v=0.5+v_r(\tau_l-0.5)$ where $v_r$ is the kinematic viscosity ratio between the vapor and liquid phases. In the next subsections, results of the impact test for different conditions are shown. 

\subsection{Impact Under Low Reynolds Number}
\label{sec:Low_Reynolds}

In this subsection, two impact tests are conducted with impact velocities of $V_d=0.075$ and $\tau_l=0.7$, which results in a Reynolds number $Re=112.5$ ($Re = V_dD/\nu_l$, where D is the droplet diameter and $\nu_l$ the liquid kinematic viscosity). The first test occurs at a reduced temperature of 0.6, and the EOS parameters are selected in order to give an interface width of 8. In the second test, the reduced temperature is 0.5 and interface width is set as equal to 7. The EOS parameters for these two cases are shown in Table~(\ref{tab:DropImpactParameters}), also with the surface tension and vapor density relative errors for the planar interface.
Differently from the static tests where the Peng bulk phases was equal to the C-S EOS with $a=0.5$, now
the Peng EOS was constructed to match the bulk phases of the C-S EOS with the $a$ values used in Table~(\ref{tab:DropImpactParameters}).
In this way, when $Tr=0.6$ the Peng EOS bulk phases correspond to the C-S EOS with $a=0.387$ and for $Tr=0.5$ we have bulk phases equal to C-S EOS with $a=0.363$. We are matching all bulk phases to avoid compressibility differences between simulations. The same approach is adopted for the Smooth EOS liquid phase. The only difference is that the vapor phase is modelled by the C-S EOS with $a=2$. We opted to increase the vapor region sound speed because, as it was seen in Fig.~(\ref{fig:Saturation_Numerical_Smooth_high_slope}), this procedure reduces the vapor phase density error.  

%
\begin{table}[htbp]
\centering
\caption{Equation of state parameters ($r_{\theta}$, $a$ and $\alpha$), surface tension ($\gamma_{\theta}$, $\gamma_{a}$ and $\gamma_{\alpha}$) and vapor density error ($E_{\theta}$, $E_{a}$ and $E_{\alpha}$) for the Peng C-S, C-S and Smooth C-S EOS respectively, at $T_r = 0.6$, $0.5$ and interface width $w=8$, $7$.
\label{tab:DropImpactParameters} }
\begin{ruledtabular}
\begin{tabular}{ccc}
\textrm{\phantom{L}}&
\textrm{$T_r = 0.6 ~ (w = 8)$}&
\textrm{$T_r = 0.5 ~ (w = 7)$} \\
\hline
$r_{\theta}$ & 0.45 & 0.44 \\
$\gamma_{\theta} \cdot 10^{3}$ & 6.5 & 8.8 \\
$E_{\theta} (\%)$ & 0.0046 & 0.10 \\
\hline
$a$ & 0.387 & 0.363 \\
$\gamma_a \cdot 10^{3}$ & 6.5 & 8.9 \\
$E_{a} (\%)$ & 5.88 & 35.67 \\
\hline
$\alpha$ & 0.834 & 0.775 \\
$\gamma_{\alpha} \cdot 10^{3}$ & 6.3 & 8.3 \\
$E_{\alpha} (\%)$ & 1.49 & 8.16 \\
\end{tabular}
\end{ruledtabular}
\end{table}

The surface tension values shown in Table~(\ref{tab:DropImpactParameters}) were computed by running a planar interface simulation.
Then the surface tension is calculated using the Eq.~(\ref{eq:IntegralSurfaceTension}).
But, the right-hand side of this equation was manipulated using $\partial_x(\psi\partial_x\psi) = \psi \partial_x\partial_x\psi + (\partial_x\psi)^2$, which resulted in the following relation:
\begin{equation}
\label{eq:SurfaceTension}
    \gamma = - \frac{Gc^4}{6} \int_{-\infty}^{\infty} \left( \frac{d\psi}{dx} \right)^2 dx.
\end{equation}
This equation was solved using a fourth order centered discretization for the spatial derivative of $\psi$ and the Simpson's rule for the numerical integration.
Since the surface tension is slightly different for each method, the values $V_d=0.075$ and $\tau_l=0.7$ are fixed for the C-S EOS. In order to maintain the same Weber number $We=\rho_lV_d^2D/\gamma$ and Reynolds number, the impact velocity $V_d$ and relaxation time $\tau_l$ are adjusted in simulations with the Peng and Smooth C-S EOS.
 
\begin{figure}[H]
\centering
	\includegraphics[width=\columnwidth]{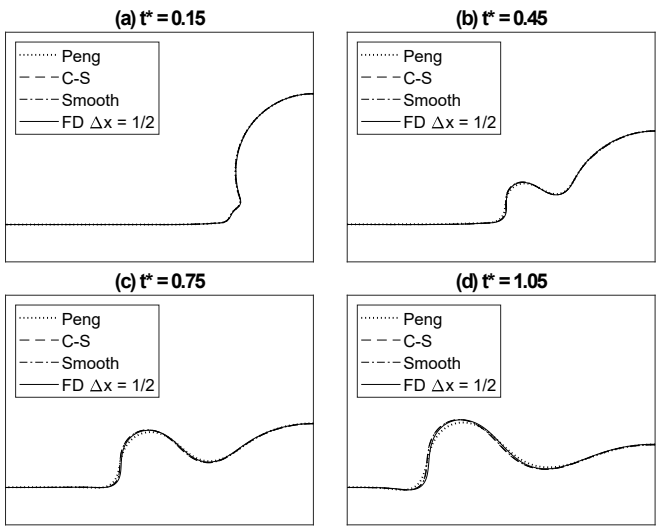}
	\caption{Snapshots of a droplet impact on a thin liquid film for different dimensionless time $t^{\ast}=tU/D$ using different methods for $T_r = 0.6$, $v_r=5$ and $Re = 112.5$.}
	\label{fig:Comp_High_Tr_Low_Re}
\end{figure}

The results for the first test are shown in Fig.~(\ref{fig:Comp_High_Tr_Low_Re}). We can see that all EOS provided good results, close to the FD solution. The viscosity ratio in this simulation was set as $v_r=5$, this option was made because the Peng method showed to be unstable for lower values of viscosity ratio. 
For a reduced temperature of $T_r=0.6$, the Maxwell rule applied to the C-S EOS (with $b=4$ and $R=1$) gives a density ratio close to 132. 

\begin{figure}[H]
\centering
	\includegraphics[width=\columnwidth]{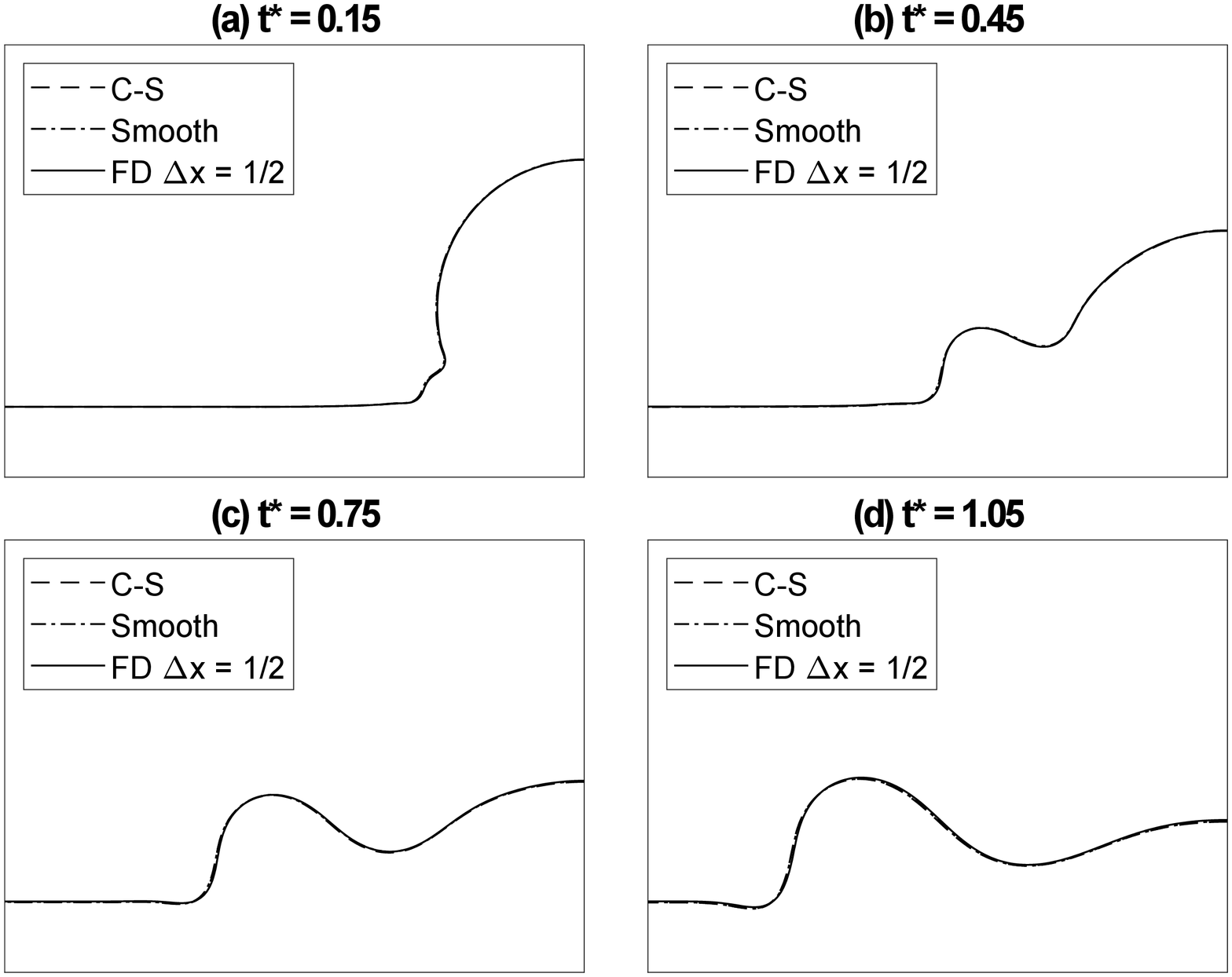}
	\caption{Snapshots of a droplet impact on a thin liquid film for different dimensionless time $t^{\ast}=tU/D$ using different methods for $T_r = 0.5$, $v_r=1$ and $Re = 112.5$.}
	\label{fig:Comp_low_Tr_Low_Re}
\end{figure}

The results of the test done at $T_r=0.5$ are shown in Fig.~(\ref{fig:Comp_low_Tr_Low_Re}). 
For this reduced temperature, the Maxwell rule applied to the C-S EOS (with $b=4$ and $R=1$) gives a density ratio close to 724. In the simulation with the Smooth C-S EOS (using $\alpha=0.775$) the numerical density ratio is 700 and for the original C-S EOS (using $a=0.363$) the density ratio is 534.  
In Fig.~(\ref{fig:Comp_low_Tr_Low_Re})
it was observed that the Peng method does not provide a stable simulation for any value of $v_r$ at this reduced temperature. In this way, the comparison between the C-S and Smooth C-S EOS are done for $v_r=1$. Both methods provided good results in the dynamic test even for a high value of density ratio. 

It was expected that the vapor phase would have a small impact in this test due to the differences in inertia and viscosity between the two phases. But the good results obtained with the C-S EOS depiste the large errors in the vapor density, suggest us that the discretization errors that generated the large discrepancy in the vapor density does not affected the overall dynamic behaviour of the method. 


\subsection{Impact Under High Reynolds Number}
\label{sec:High_Reynolds}

\begin{figure}
\centering
	\includegraphics[width=\columnwidth]{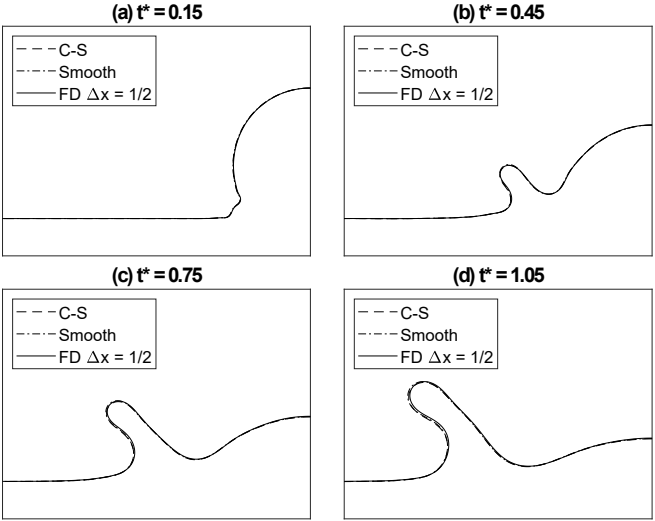}
	\caption{Snapshots of a droplet impact on a thin liquid film for different dimensionless time $t^{\ast}=tU/D$ using different methods for $T_r = 0.6$, $v_r=1$ and $Re = 300$.}
	\label{fig:Comp_High_Tr_High_Re}
\end{figure}

In this subsection, the impact tests are conducted under higher Reynolds number conditions than the previous tests. The droplet impact velocity is increased to $V_d=0.1$ and the relaxation time is reduced to $\tau_l=0.6$, resulting in $Re=300$. Again, two simulation conditions are selected following the parameters of Table~(\ref{tab:DropImpactParameters}). The parameters $V_d$ and $\tau_l$ are adjusted in the simulations with the Peng and Smooth EOS in order to maintain the same Reynolds and Weber numbers. 
For the simulation done at $T_r=0.6$ it was observed that the Peng method only results in a stable simulation for $v_r=15$. Since this value of $v_r$ is out of the stability region of the FD method, in this subsection it is shown only the comparison for $v_r=1$. 
Results for this case are shown in Fig.~(\ref{fig:Comp_High_Tr_High_Re}), good agreement between the different EOS was observed.
In the next subsection it is shown a comparison between the C-S, Peng and Smooth EOS for $v_r=15$.   

\begin{figure}
\centering
	\includegraphics[width=\columnwidth]{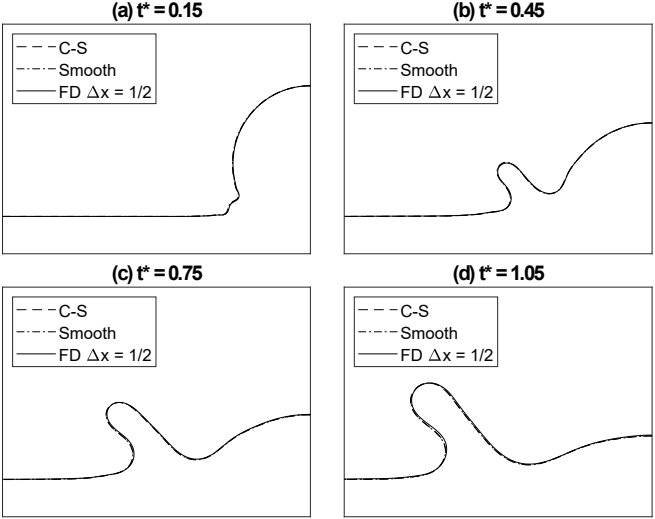}
	\caption{Snapshots of a droplet impact on a thin liquid film for different dimensionless time $t^{\ast}=tU/D$ using different methods for $T_r = 0.5$, $v_r=1$ and $Re = 300$.}
	\label{fig:Comp_low_Tr_High_Re}
\end{figure}

For the reduced temperature $T_r=0.5$ the Peng method does not resulted in stable simulation in the evaluated range of $v_r$ between 1 and 24. 
In this way, the comparison was done only for the C-S and Smooth C-S EOS setting $v_r=1$. Results for $T_r=0.5$ are shown in 
Fig.~ (\ref{fig:Comp_low_Tr_High_Re}). As in the previous cases under lower Reynolds number condition, both methods provided similar results, close to the FD method simulation. The C-S and Smooth C-S EOS showed good stability properties allied to satisfactory numerical accuracy in the dynamic tests. 

\subsection{Comments about the Forcing Scheme}

\begin{figure}
\centering
	\includegraphics[width=\columnwidth]{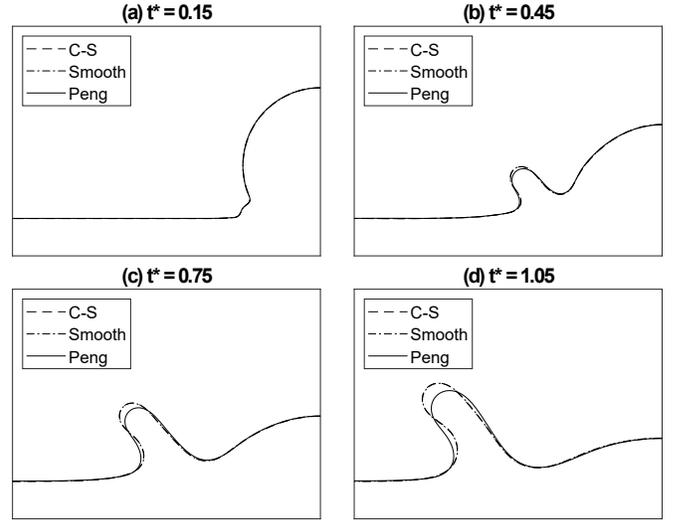}
	\caption{Snapshots of a droplet impact on a thin liquid film for different dimensionless time $t^{\ast}=tU/D$ using C-S, Smooth and Peng EOS for $T_r = 0.6$, $v_r=15$ and $Re = 300$.}
	\label{Comp_Compressibilidade_S_N_P}
\end{figure}

In the previous subsection, for the simulation at $T_r=0.6$ and $Re=300$, the Peng method was not included in the comparisons because it was not stable in the viscosity ratio $v_r$ stability range of the FD method. In this subsection we bring a direct comparison between the C-S, Smooth and Peng EOS using $v_r=15$. The results are shown in Fig.~(\ref{Comp_Compressibilidade_S_N_P}). It is noted a discrepancy between the Peng EOS with the other methods. Great agreement between the C-S and Smooth EOS was obtained. 

\begin{figure}
\centering
	\includegraphics[width=\columnwidth]{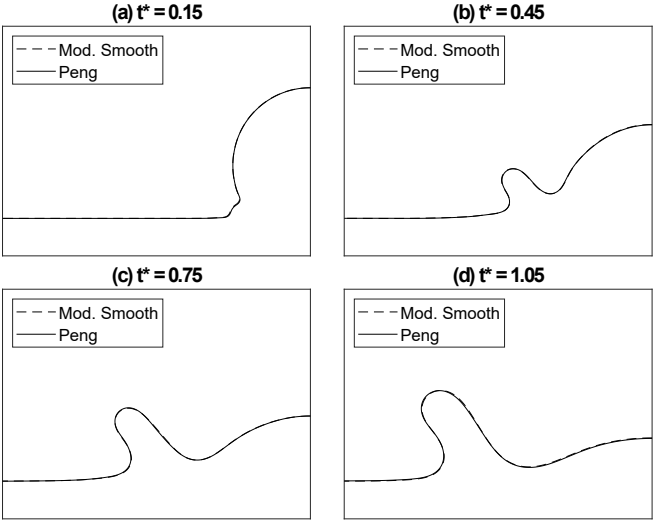}
	\caption{Snapshots of a droplet impact on a thin liquid film for different dimensionless time $t^{\ast}=tU/D$ using Modified Smooth and Peng EOS for $T_r = 0.6$, $v_r=15$ and $Re = 300$.}
	\label{Comp_Compressibilidade_Nm_P}
\end{figure}

Next we investigate the reasons for this difference and if it is related with the EOS. Both C-S and Smooth EOS are implemented using the 
\citet{li2013lattice} forcing scheme. The Peng method is implemented directly using the \citet{guo2002discrete} forcing scheme. Since in this last case the parameter $\epsilon$ of Eq.~(\ref{eq:VaporLiqEq}) can not be controled to ajdust the thermodynamic consistency, the parameter $\rho_m$ of Eq.~(\ref{eq:PengEOS}) is computed numerically to made the EOS satisfy the mechanical stability condition with $\epsilon=0$ for the desired vapor and liquid densities. In the Smooth EOS, the parameter $P_{max}$ is computed in order to satisfy the Maxwell rule, but we can also change this condition and compute $P_{max}$ to satisfy the mechanical stability condition for $\epsilon=0$. In this way, we create a modified Smooth EOS which can be implemented directly with the \citet{guo2002discrete} forcing scheme in the same way as the Peng method. We called this EOS as Modified Smooth and it was computed using $\alpha=0.446$ at $T_r=0.6$. The new EOS result in an interface width of $w=8$ with $\gamma=0.0060$. Using this $\alpha$ parameter to set the Mod. Smooth we can perform a comparison with the Peng EOS in the same conditions of the simulations shown in Fig.~(\ref{Comp_Compressibilidade_S_N_P}). The results of this comparison are shown in Fig.~(\ref{Comp_Compressibilidade_Nm_P}). In this test, excellent agreement was obtained between simulations even both use different EOS. These results suggest that the differences observed in Fig.~(\ref{Comp_Compressibilidade_S_N_P}) are mainly due to the different forcing schemes and not due to the different EOS. Since the focus of this work is on the EOS impact on the simulation results we will let the study of this forcing scheme influence as a suggestion for future works.

\begin{figure}[H]
\centering
	\includegraphics[width=\columnwidth]{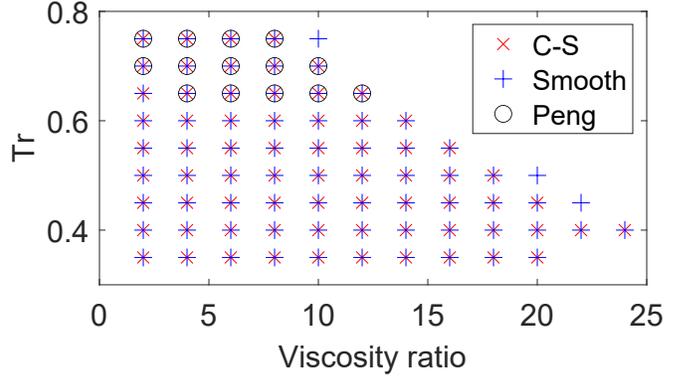}
	\caption{Stability map for different EOS in respect to the reduced temperatue $T_r$ and viscosity ratio $v_r$.
	}
	\label{fig:StabilityMap}
\end{figure}

In order to evaluate in more detail the stability properties of the different EOS, several simulations are conducted at different simulation conditions. We perform simulations in a range of reduced temperatures between $T_r=0.35$ and $T_r=0.75$.
We fixed $a=0.363$ for the C-S EOS, and the parameters for the other equations of state were selected in such a way to maintain the same interface width (at each temperature) for all methods.
So, for each temperature, we measure the interface width (in a planar interface test) of the method that use the C-S EOS. Then, we set the parameters of the other equations of state that gives the same interface width for each temperature tested.
The viscosity ratio $v_r$ is also varied in a range between 2 to 24 in steps of 2. All simulations are done for $V_d=0.1$ and $\tau_l=0.6$ which give us $Re=300$. A map of stability is shown in Fig.~(\ref{fig:StabilityMap}) in which each point means that the simulation was stable for that values of $T_r$ and $v_r$. 
We can see that simulations using the Peng C-S EOS showed reduced stability for low values of reduced temperature. On the other hand, the Smooth C-S EOS maintained the method stability until a reduced temperature of $T_r=0.35$.  
The density ratio reached by the simulation with the Smooth C-S at this temperature was approximately $3.38\cdot10^{4}$ while the original C-S simulation gives $8.85\cdot10^{3}$. The density ratio given by the thermodynamic consistent phase densities is $7.85\cdot10^{4}$.
In Fig.~(\ref{fig:StabilityMap}) the stability region of the Peng method does not cover the reduced temperature $T_r=0.6$, but in the previous simulation we were able to obtain stable results for $v_r=15$. It should be considered that this stability map is valid for the specific set of parameters used in these simulations. In this case we set the parameters of all methods in order to have the same interface width as the C-S EOS with $a=0.363$, at this condition the maximum density ratio obtained with the Peng method was 68 at $T_r=0.65$. At different conditions we can have other behaviours in terms of stability.

After the dynamic study performed in this section, the following conclusions can be made. 
The dynamic tests revealed that the Peng method, despite its better static test results, suffers with instability issues for
high values of density ratio and Reynolds number. In the tests performed along this section, simulations with the C-S EOS provided excellent agreement
with the FD reference solution. This fact suggest that discretization errors that were largely deviating the vapor density values does not
affect the overall dynamic behaviour of the method. It should be noted that in tests done in this section, the vapor region have a small 
impact due to the high difference in inertia and viscosity (dynamic viscosity) between phases. In a phase-change problem, where heat is a 
absorbed to convert liquid into vapor, this large deviation in the vapor density would have
a great impact on simulation results. The vapor density deviation can be mitigated by using the Smooth C-S to increase the vapor region slope. 
This customized EOS also showed very good stability under high values of density ratio and Reynolds number in the dynamic tests. Excellent agreement was obtained by comparing LBM simulation results for the Smooth C-S with the FDM simulation results.  

Another limitation related with the standard procedure, is that by multiplying the equation of state by a small factor, the EOS sound speed is reduced. This procedure affects the simulation Mach number which can lead to the influence of compressibility effects in the simulation. In order to correct this issue it would be necessary to reduce the impact velocity to maintain the Mach number and re-scale the problem 
to achieve the same dimensionless numbers. In this way, due to the smaller velocity the simulation computational time would be increased. 
The Smooth C-S does not have this limitation because the vapor and liquid region slope are not changed when the $\alpha$ parameter is modified.
When the $\alpha$ parameter is reduced, the surface tension is also reduced which affects the Weber number. In this work the impact velocity was used to control the Weber number, but a surface tension controlling technique \cite{li2013achieving} could be use to avoid any change in the droplet impact velocity. 


\section{Conclusion}
\label{sec:conclusion}

In this work, we studied a different approach to enhance the pseudopotential method accuracy and stability that does not involve an increase in the complexity of the numerical scheme and does not require higher order approximations. Instead we follow a recently proposed approach \cite{peng2020attainment} of altering the vdW loop in the EOS.
In our simulations we found that the Peng \textit{et al.} \cite{peng2020attainment} approach gave excellent results in equilibrium tests but was less stable in the dynamic tests, particularly at large density ratios. 

Motivated by this fact, we introduced a new procedure to change the EOS vdW loop which, differently from the previous work done by Peng \textit{et al.} \cite{peng2020attainment}, maintains the continuity of the sound speed in the stable and meta-stable branches and also respects the Maxwell equal area rule. We conducted static planar interface tests (using the C-S EOS) and compared simulations using the standard procedure of multiplying the EOS by a small factor, the vdW loop replacement proposed by Peng \textit{et al.} \cite{peng2020attainment} and the novel procedure that we proposed in this work. We observed that in general when the EOS is modified in such a way that the interface thickness is increased, the numerical results approaches the ones predicted by the mechanical stability condition.

The relative error of the vapor density (in comparison with the thermodynamic consistent vapor density) decreases from 35\% to 12\% when interface thickness is increased from 7 to 11 lattice units for a reduced temperature of $T_r=0.5$. For this same temperature, when the currently proposed vdW loop modification is used, the relative error of the vapor density is 8\% for an interface thickness of 7 lattice units and this error decreases below 2\% when interface thickness is increased to 11 lattice units. The Peng \textit{et al.} \cite{peng2020attainment} method provided excellent static test results with errors below 0.15\% for all tested interface widths.

Since no analytical solutions for the impact of a droplet exist, it can be challenging to evaluate the accuracy of each method in the dynamic tests using this benchmark problem. For comparison we developed a novel Finite Difference scheme that is exactly matched to the hydrodynamic equations predicted for the lattice Boltzmann methods. The Finite Difference scheme refinement is trivial, and therefore allows us to obtain a reference solution against which we can compare the other methods. 

In the dynamic tests done in this work of a droplet impact on a liquid film with Reynolds number of 300, the minimum reduced temperature achieved by the Peng method was $T_r=0.65$ and a density ratio up to 68. The currently proposed procedure allowed simulations at the same condition with a reduced temperature of $T_r=0.35$ and density ratio up to $3.38\cdot10^{4}$. Excellent agreement was obtained by simulations using our procedure  against the reference finite difference solution of the droplet impact problem, done for a reduced temperature of $T_r=0.5$. These results showed that the proposed procedure is accurate both for equilibrium and non-equilibrium simulation while allows simulations with high density ratios.

\section*{Acknowledgments}
\label{appendix-sec1}

The authors acknowledge the support received from 
CAPES (Coordination for the Improvement of Higher Education Personnel, Finance Code 001), 
from CNPq (National Council for Scientific and Technological Development, processes 431782/2018-0 and 140634/2019-3) and 
FAPESP (São Paulo Foundation for Research Support, 2016/09509-1 and 2018/09041-5), 
for developing research that have contributed to this study.

\appendix

\section{Finite Difference Method}
\label{sec:AppendixFDmethod}

In this work, the goal is to evaluate the impact of equation of state modifications in the pseudopotential results
in static and dynamic tests. For some dynamic tests it is difficult to obtain or it is not known analytical solutions. 
For this reason, in this work it was developed a finite difference (FD) method that replicates the solution of the 
pseudopotential LBM. With this FD method it is possible to perform a mesh refinement and obtain reference solutions for
the cases in which we want to evaluate the LBM solution. 
The mass and momentum conservation equations are written in the following form:
\begin{subequations}
\begin{equation}
    \partial_t \rho = - \left( \rho \partial_{\gamma} u_{\gamma} + u_{\gamma} \partial_{\gamma} \rho \right),
\end{equation}
\begin{equation}
\label{eq:FDmomentum}
    \partial_t u_{\alpha} = - u_{\gamma} \partial_{\gamma} u_{\alpha} 
    - \frac{1}{\rho} \partial_{\alpha} p_{EOS} - \frac{1}{\rho} \partial_{\beta} p'_{\alpha\beta}
    + \frac{1}{\rho} \partial_{\beta} \sigma'_{\alpha\beta},
\end{equation}
\end{subequations}
where the pressure tensor $p_{\alpha\beta}$ was divided in two components $p_{EOS}\delta_{\alpha\beta}$ and $p'_{\alpha\beta}$.
These equations are discretized following the predictor-corrector approach used in the MacComark's method \cite{anderson1995computational}.
In the mass conservation equation the terms $\partial_{\gamma}u_{\gamma}$ and $\partial_{\gamma}\rho$ are discretized using first order forward differences in the predictor step and backward stencils in the corrector step. 
The same procedure is applied to the terms $\partial_{\gamma}u_{\alpha}$ and $\partial_{\alpha}p_{EOS}$ in the momentum conservation
equation. The viscous stress tensor $\sigma'_{\alpha\beta}$ is given by Eq.~(\ref{eq:StressTensor}) with the dynamic viscosities given by Eq.~(\ref{eq:DynamicViscosity}).
Then we write the term $\partial_{\beta}\sigma'_{\alpha\beta}$ in Eq.~(\ref{eq:FDmomentum}) as:
\begin{align}
\label{eq:FDstressTensor}
    \partial_{\beta} \sigma'_{\alpha\beta} & = (\partial_{\beta} \mu) \left( \partial_{\alpha} u_{\beta} + \partial_{\beta} u_{\alpha} \right) + \mu \left( \partial_{\alpha} \partial_{\beta} u_{\beta} + \partial_{\beta} \partial_{\beta} u_{\alpha} \right) \\ \nonumber
    & + ( \partial_{\alpha} \mu_B ) \partial_{\gamma} u_{\gamma} 
    + \mu_B \partial_{\alpha} \partial_{\beta} u_{\beta} , 
\end{align}
the dynamic viscosities Eq.~(\ref{eq:DynamicViscosity}) depends on the density and the relaxation times that are also functions of the density, Eq.~(\ref{eq:RelaxationTime}). Then, we can write the terms $\partial_{\beta}\mu$ and $\partial_{\alpha}\mu_B$ as:
\begin{equation}
\label{eq:FDviscosityDerivative}
    \partial_{\beta} \mu = \left( \partial_{\rho} \mu \right) \left( \partial_{\beta} \rho \right) ~~~~~
    \partial_{\alpha} \mu_B = \left( \partial_{\rho} \mu_B \right) \left( \partial_{\alpha} \rho \right),
\end{equation}
the first order derivatives of Eqs.~(\ref{eq:FDstressTensor}) and (\ref{eq:FDviscosityDerivative}) are discretized using first order forward differences in the predictor step and backward differences in the corrector step. The second order derivatives are discretized using second order central schemes for both the predictor and corrector step. The pressure tensor component $p'_{\alpha\beta}$ can be obtained from the LBM pressure tensor given by Eqs.~(\ref{eq:ShanPressureTensor}) and (\ref{eq:NewPressureTensor}):
\begin{equation}
\begin{aligned}
    p'_{\alpha\beta} &= \left( -\epsilon \frac{Gc^4}{8} (\partial_{\gamma}\psi)(\partial_{\gamma}\psi) 
    + \frac{Gc^4}{12} \psi \partial_{\gamma}\partial_{\gamma} \psi \right) \delta_{\alpha\beta} \\
    & + \frac{Gc^4}{6}\psi\partial_{\alpha}\partial_{\beta}\psi,
\end{aligned}
\end{equation}
the term $\partial_{\beta}p'_{\alpha\beta}$ in Eq.~(\ref{eq:FDmomentum}) can be written as:
\begin{align}
\label{eq:FDforce}
    \partial_{\beta} p'_{\alpha\beta} & = -\epsilon \frac{Gc^4}{4} (\partial_{\gamma} \psi)(\partial_{\alpha}\partial_{\gamma}\psi)
    + \frac{Gc^4}{12}(\partial_{\alpha}\psi)(\partial_{\gamma}\partial_{\gamma}\psi) \\ \nonumber
    & + \frac{Gc^4}{6} (\partial_{\gamma}\psi)(\partial_{\alpha}\partial_{\gamma}\psi) 
    + \frac{Gc^4}{4} \psi \partial_{\alpha}\partial_{\gamma}\partial_{\gamma}\psi,
\end{align}
in this equation, all the derivatives are discretized using second order central differences for both predictor and corrector step.


\section{Grid-Dependency Study}
\label{sec:GridStudy}

In this appendix, a grid dependency study for the droplet impact test with the FD method is performed. 
An appropriate boundary condition must be defined for the FD method in order to avoid discrepancies in the comparison
with the LBM that are not due to discretization errors. In the LBM method, we used the bounce-back method to set the physical boundary
displaced of half lattice from the computational boundary. Also, for the pseudopotential force computation at the computational boundary, it was considered that the node inside the wall (bottom boundary) have the same density as the saturated liquid. 
In the FD method, we need to set two boundary layers because in Eq.~(\ref{eq:FDforce}) we are computing the derivative $\partial_{\alpha}\partial_{\gamma}\partial_{\gamma}\psi$ using a second order central difference. In the bottom boundary, we will set the first boundary layer as been infinitesimally close to the solid wall. In this way, the non-slip boundary condition is applied. 
Also, since the LBM is a pseudo-compressible method we will let the density at this boundary float by using an extrapolation from the domain density. In this way, the boundary condition at the first bottom boundary layer (b1) is:
\begin{align}
\label{eq:B1}
    \rho_{b1} & = 2 \rho_{0} - \rho_{1} \\ \nonumber
    \psi_{b1} & = 2 \psi_{0} - \psi_{1} \\ \nonumber
    U_{b1} & = 0 \\ \nonumber
    V_{b1} & = 0, \\ \nonumber
\end{align}
the second boundary layer (b2) would correspond to the node inside the solid wall. In the LBM method we considered that this node have a density equal to the liquid density. But in the LBM since we are using the bounce back, the solid node is half lattice from the physical boundary. As each lattice have a grid spacing of $\Delta y = 1$, the distance between the solid node and the physical boundary is $\Delta = 0.5$. In the FD method, the solid node (at b2) is displaced $\Delta y$ from the physical boundary and now we can change the grid spacing to any desired value. In this way we adopted the following density value the solid node to compensate the different displacement in comparison with the LBM:
\begin{equation}
\label{eq:B2}
    \psi_{b2} = \psi_{b1} + 2\Delta y ( \psi_l - \psi_{b1} ),
\end{equation}
where $\psi_l=\psi(\rho=\rho_l)$. It is not necessary to specify the other variables at b2. In the top boundary we define the last fluid node in the position N as being infinitesimally close to the wall. A simple and stable way to impose non-slip condition at this position is by defining a reflexive boundary in the top boundary layers (t1 and t2):
\begin{align}
    \rho_{t1} & = \rho_{N-1} \\ \nonumber
    \psi_{t1} & = \psi_{N-1} \\ \nonumber
    U_{t1} & = - U_{N-1} \\ \nonumber
    V_{t1} & = - V_{N-1}, \\ \nonumber
    \rho_{t2} & = \rho_{N-2}, \\ \nonumber
\end{align}
since we inverted the velocities direction from the position $N-1$ to t1 in the y-direction, it is equivalent to impose $U_N=0$ and $V_N=0$. At the side boundaries we applied periodic boundary condition. 

\begin{figure}[h]
\centering
	\includegraphics[width=80mm]{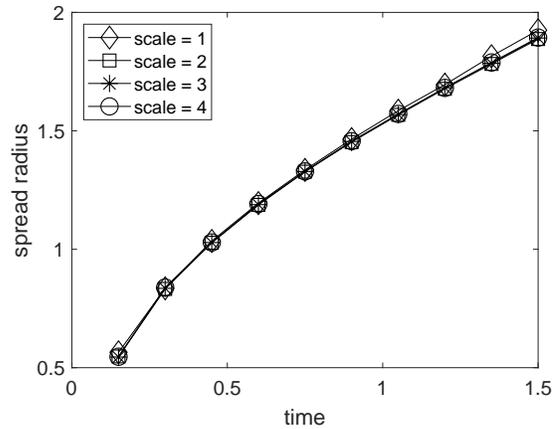}
	\caption{Comparison of the spread ratio as a function of time for different mesh sizes using the FD method with the C-S EOS.}
	\label{fig:Convegence_FD_I8}
\end{figure}

With the boundary conditions defined, we applied the FD method to simulate droplet impact in a thin liquid film problem. 
We use the same geometric parameters employed in the LBM simulation described in Sec.~(\ref{sec:DynamicResults}). The C-S EOS is used to define $p_{EOS}$ in Eq.~(\ref{eq:FDmomentum}). The EOS parameters are selected to result an interface width $w=8$ for the reduced temperature $T_r=0.6$ following the Table~(\ref{tab:DropImpactParameters}). The FD method time step was defined as $\Delta t = 0.8 (\Delta x)^2$ when $v_r=1$. It was noticed that the method becomes less stable for other values of viscosity ratios. In this way for $v_r\neq1$ it is used $\Delta t = 0.2 (\Delta x)^2$. Stable simulation results were obtained for $\Delta x = 1$ and $\Delta x = 2$. But for smaller $\Delta x$ simulations become unstable. By modifying the bottom boundary conditions to $\rho_{b1}=\rho_l$, $\psi_{b1}=\psi_l$ and $\psi_{b2}=\psi_l$ we were able to obtain stable results for different meshes. Following the definition of spread ratio $r_s$ used by Josserand and Zaleski \cite{josserand2003droplet}, we show the results for the dimensionless spread ratio $r^{\ast}=r_s/D$ as a function of the dimensionless time $t^{\ast}=tU/D$ in Fig.~{\ref{fig:Convegence_FD_I8}}. The viscosity ratio used in these simulations was $v_r=1$.

We can see that the mesh with $\Delta x = 1/2$ provided very close results with the finer mesh of $\Delta x=1/4$. 
The boundary condition used in this test provided stable results for different mesh sizes, but it was noted that the imposition of the 
liquid density at the boundary $\rho_{b1}=\rho_l$ can affect the results depending on the simulation Mach number. The reason is that the LBM is a pseudo-compressible method, so the densities at the boundary can slightly float due to pressure changes. Fix the bottom density is also equivalent to fix the pressure. But for the tests done in this work we found that the boundary condition established in Eqs.~(\ref{eq:B1}) and (\ref{eq:B2}) are in physical agreement with the LBM simulation and will be adopted along this work. In this case, the finest mesh that can be used is $\Delta x = 1/2$, but results shown in Fig.~(\ref{fig:Convegence_FD_I8}) suggest that this mesh provide accurate results. 

\providecommand{\noopsort}[1]{}\providecommand{\singleletter}[1]{#1}%

\end{document}